\documentclass{aa}
\usepackage{graphicx}
\usepackage{txfonts}
\begin{document}

\title{Synthetic photometry for carbon rich giants}
\subtitle{I. Hydrostatic dust-free models}

\author{B.~Aringer\inst{1,2,3} \and L.~Girardi\inst{2} \and
W.~Nowotny\inst{1} \and P.~Marigo\inst{3} \and
M.T.~Lederer\inst{1}}

\institute{Department of Astronomy, University of Vienna,
T\"urkenschanzstra{\ss}e 17, A-1180 Wien, Austria\\
\email{aringer@astro.univie.ac.at}
\and
Osservatorio Astronomico di Padova -- INAF,
Vicolo dell'Osservatorio 5, I-35122 Padova, Italy\\
\email{leo.girardi@oapd.inaf.it}
\and
Dipartimento di Astronomia, Universit\`a di Padova,
Vicolo dell'Osservatorio 2, I-35122 Padova, Italy\\
\email{paola.marigo@unipd.it}}

\date{Received; accepted}

\titlerunning{Synthetic photometry for carbon-rich giants}
\authorrunning{B. Aringer et al.}

\abstract
{Carbon rich objects represent an important phase during the late stages of
evolution of low and intermediate mass stars. They contribute significantly
to the chemical enrichment and to the infrared light of galaxies. A proper
description of their atmospheres is crucial for the determination of
fundamental parameters such as effective temperature or mass loss rate.}
{We study the spectroscopic and photometric properties of carbon
stars. In the first paper of this series we focus on objects that can
be described by hydrostatic models neglecting dynamical phenomena
like pulsation and mass loss. As a consequence, the reddening
due to circumstellar dust is not included. Our results are collected
in a database, which can be used in conjunction with stellar evolution
and population synthesis calculations involving the AGB.}
{We have computed a grid of 746 spherically symmetric COMARCS
atmospheres covering effective temperatures between 2400 and
4000~K, surface gravities from $\rm log(g~[cm/s^2]) = 0.0$ to
$-1.0$, metallicities ranging from the solar value down to
one tenth of it and C/O ratios in the interval between 1.05
and 5.0. Subsequently, we used these models to create synthetic
low resolution spectra and photometric data for a large number
of filter systems. The tables including the results are
electronically available. First tests of the application
on stellar evolution calculations are shown.}
{We have selected some of the most commonly used colours in order
to discuss their behaviour as a function of the stellar parameters.
A comparison with measured data shows that down to 2800~K the
agreement between predictions and observations of carbon
stars is good and our results may be used to determine quantities
like the effective temperature. Below this limit the synthetic
colours are much too blue. The obvious reason for these problems
is the neglect of circumstellar reddening and structural changes
due to pulsation and mass loss.}
{The warmer carbon stars with weak pulsation can be
successfully described by our hydrostatic models. In order
to include also the cooler objects with intense variations,
at least a proper treatment of the reddening caused by the
dusty envelopes is needed. This will be the topic of the
second paper of this series.}


\maketitle

\section{Introduction}

During the evolution of low to intermediate mass stars along the
TP-AGB (Thermal Pulsing Asymptotic Giant Branch), material processed by the
nuclear reactions in the He burning shell may be dredged up to the surface,
changing the chemical abundances in the atmosphere. Especially, the amount
of carbon will be increased significantly by this mechanism.
If the particle density exceeds that of oxygen (the ratio of
the particle densities $\rm C/O > 1$), the spectral
appearance of the object becomes completely different. Instead
of O-bearing molecules like TiO, VO, SiO, OH and H$_2$O the opacities
in the cool outer layers and the observed energy distributions
are dominated by C$_2$, CN, C$_3$, HCN and C$_2$H$_2$ (e.g.\
Querci et al.~\cite{cphotque74}, J{\o}rgensen et al.~\cite{cphotjor00},
Loidl et al.~\cite{cphotloi01}). This characterizes a classical
carbon star.

Carbon stars are among the brightest stellar objects in
resolved galaxies containing young and intermediate-age
populations, especially in the near infrared. Moreover, they
contribute significantly to the integrated spectra of such systems.
These two facts become obvious in the Magellanic
Clouds: in the LMC for instance, among the approximately 31000 AGB
stars brighter than the tip of the RGB (Cioni \& Habing~\cite{cphotcio03}),
there are about 11000 carbon rich objects (Blanco \& McCarthy~\cite{cphotbla83}).
These sources alone contribute to roughly 20 percent of the integrated
bolometric luminosity of LMC clusters with intermediate ages
(Frogel et al.~\cite{cphotfro90}, Fig.~16). Thus, the influence of C-type giants
on the total flux emitted by galaxies is noticeable.

It is therefore very important that carbon stars are properly taken into
account in models of galaxies. For this, two
major ingredients are necessary. First, evolutionary tracks providing the
distributions of luminosities, effective temperatures and surface
compositions of the red giants as a function of age, initial mass and
metallicity have to be available. Such calculations were presented
for example in the work of Marigo \& Girardi~(\cite{cphotmar07}).
Secondly, one needs synthetic spectra including the circumstellar
reddening by dust, which allow the conversion of the model quantities
into observable properties of stars. The implementation of this
can be found in Marigo et al.~(\cite{cphotmar08}).

The main goal of this work is to present a set of photometric
data and low resolution energy distributions allowing such a
conversion from stellar fundamental parameters into measurable
quantities. Based on observational material, libraries of overall
carbon star spectra have been published by Lan\c{c}on \&
Wood~(\cite{cphotlan00}) and Lan\c{c}on \& Mouhcine~(\cite{cphotlan02}).
However, they extend only up to 2.5~$\mu$m, neglecting the mid and
far infrared. In addition, their coverage of stellar parameters
is very limited. This applies also to the input data used by
Marigo et al.~(\cite{cphotmar08}) which are based on the hydrostatic
model atmospheres and synthetic spectra of Loidl et al.~(\cite{cphotloi01}).
As an example, they only have included calculations for solar metallicity.

Loidl et al.~(\cite{cphotloi01}) and J{\o}rgensen et al.~(\cite{cphotjor00})
have computed synthetic low and medium resolution spectra based on
hydrostatic carbon star models and compared them to observations in the optical
as well as in the near and mid infrared range. Their results were
obtained using previous versions of COMA and the MARCS code
(see Sect.~2.1). One of the most important differences to
the work presented here is that they have completely neglected
atomic line opacities, which causes flux deviations mainly at shorter
wavelengths. In addition, there are also changes concerning the
molecular input data. A comparison of the results based on our
calculations to convert the isochrones of Marigo et al.~(\cite{cphotmar08})
to the 2MASS $\rm M_{K_s}$ versus $\rm (J-K_s)$ diagram and the original
ones obtained with the spectra from Loidl et al.~(\cite{cphotloi01})
is shown in Sect.~4.2.2 (Fig.~\ref{aricphot19}).

It was demonstrated for example by Gautschy-Loidl et al.~(\cite{cphotgau04}),
Loidl et al.~(\cite{cphotloi99}) or Nowotny et
al.~(\cite{cphotnoa05}, \cite{cphotnob05}) that hydrostatic models
do not reproduce the atmospheric structure and spectra
of cool carbon stars with intense variations and mass loss.
Such objects are dominated by their pulsation creating shock
waves, the formation of dust and a strong wind giving rise to a
circumstellar shell. Since all of these phenomena are time-dependent
and coupled, they have to be described by comprehensive
dynamical calculations, such as presented by
H\"ofner et al.~(\cite{cphothof03}) or Mattsson et al.~(\cite{cphotmat08}).
The photometric properties of the corresponding models will be
discussed in the next paper of this series. In the current
work we focus on warmer objects with small pulsation amplitudes.
Nevertheless, the reddening due to circumstellar dust, which causes
the most important differences between the presented hydrostatic
and dynamical atmospheres for the cool mass-losing carbon stars,
may be simulated in connection with our data by using approximative
approaches such as in Bressan et al.~(\cite{cphotbre98}) and
Groenewegen~(\cite{cphotgro06}).

\section{Model atmospheres and spectral synthesis}
\subsection{Hydrostatic models and opacities}

In order to study the spectra and photometric properties of carbon
stars, we have produced a grid of 746 spherically symmetric hydrostatic
model atmospheres. They were computed with the COMARCS program
which is based on the version of the MARCS code (Gustafsson et
al.~\cite{cphotgus75}, Gustafsson et al.~\cite{cphotgus08}) used in
J{\o}rgensen et al.~(\cite{cphotjor92}) and
Aringer et al.~(\cite{cphotari97}).
In contrast to the previous approaches, COMARCS works with
external opacity tables including the coefficients for a
two-dimensional spline interpolation in pressure and
temperature at the desired wavelengths. The applied
method is an algorithm developed by
Hardy~(\cite{cphothar71})
with the implementation found in the book of
Sp\"ath~(\cite{cphotspa91}).
The frequency grid adopted for the
opacity sampling in our calculations corresponds to the one
used by Aringer et al.~(\cite{cphotari97}).
It consists of 5364 points. Such tables with spline
coefficients for the different wavelengths are
produced for each combination of microturbulence
and chemical abundances to be investigated.
This was done using the opacity generation code
COMA (Copenhagen Opacities for Model Atmospheres).

The COMA program was originally developed to compute wavelength
dependent absorption coefficients for dynamical model atmospheres
of cool giants as they are used for example in the work of
H\"ofner et al.~(\cite{cphothof03}). In addition, it was adapted
to determine weighted mean opacities for different
applications like stellar evolution calculations
(e.g.\ Cristallo et al.~\cite{cphotcri07},
Lederer \& Aringer~\cite{cphotled08}).
Starting from a given temperature-pressure or
temperature-density structure (table with values) the abundances
of many important species are evaluated based on the
equilibrium for ionisation and molecule formation.
Thus, we assume chemical equilibrium except for the
treatment of dust (and also some tests with results from
non-equilibrium calculations). As a next step, the computed
partial pressures of the neutral atoms, ions and molecules
can be used to determine the continuous and line
opacity at each selected frequency point (in LTE).
Since the first version of COMA described by
Aringer~(\cite{cphotari00}),
a considerable number of updates have been made
including a more complete treatment of ionisation
(Gorfer~\cite{cphotgor05}),
additional chemistry routines
(Gibbs energy minimisation, Falkesgaard~\cite{cphotfal01}),
the addition of atomic transitions from the VALD database
(Kupka et al.~\cite{cphotkup00})
or the possibility to take the effect and opacity of dust
into account (Nowotny et al.~\cite{cphotnow07}).
And, of course the number of incorporated molecules contributing
to the total absorption has been increased significantly.
The calculations presented here cover the following
species: CO, CH, C$_2$, SiO, CN, TiO, H$_2$O, C$_2$H$_2$,
HCN, C$_3$, OH, VO, CO$_2$, SO$_2$, HF, HCl, FeH, CrH,
ZrO and YO\@.

A table containing most of the line lists used
and the corresponding references can be found
in Cristallo et al.~(\cite{cphotcri07}).
Four of the molecules included in our work
do not appear in the mentioned collection of data:
ZrO from Plez et al.~(\cite{cphotple03}),
YO from John Littleton (priv.\ comm.),
FeH from Dulick et al.~(\cite{cphotdul03}) and CrH from
Bauschlicher et al.~(\cite{cphotbau01}). The abundance of
FeH originally was not computed in any of the chemical routines
of COMA\@. Thus, we had to add the corresponding equilibrium
constants to the program. They were determined using
the partition function for FeH given by
Dulick et al.~(\cite{cphotdul03})
and for the atomic contributors Fe and H by
Irwin~(\cite{cphotirw81}). For some of the molecules
already appearing in Cristallo et al.~(\cite{cphotcri07})
the line data changed: In the case of SO$_2$ and HCl
they have been updated from HITRAN 1996 to HITRAN 2004
(Rothman et al.~\cite{cphotrot05})
which is here now also used for OH\@. The opacity of
CO$_2$ is calculated from HITEMP
(Rothman et al.~\cite{cphotrot95})
instead of HITRAN which results in a much larger number
of transitions (1032269). Finally, the lines of HF
are taken from the list of R.H. Tipping (priv.\ comm.)
discussed in Uttenthaler et al.~(\cite{cphotutt08}).
However, all of these changes do not have a significant
effect on the atmospheric structures, since the molecules
involved are only of minor importance for the overall
absorption in the temperature range covered by the
hydrostatic models. Also the uncertainties of the
opacities from TiO and H$_2$O, where the existing
line lists show considerable differences in some spectral
regions (see e.g.\ Aringer et al.~\cite{cphotari08}),
do not influence our results. The partial pressures
of these species always remain very low in a carbon rich
environment -- at least as long as one assumes chemical
equilibrium.

\begin{figure*}
\sidecaption
\includegraphics[width=12cm,clip]{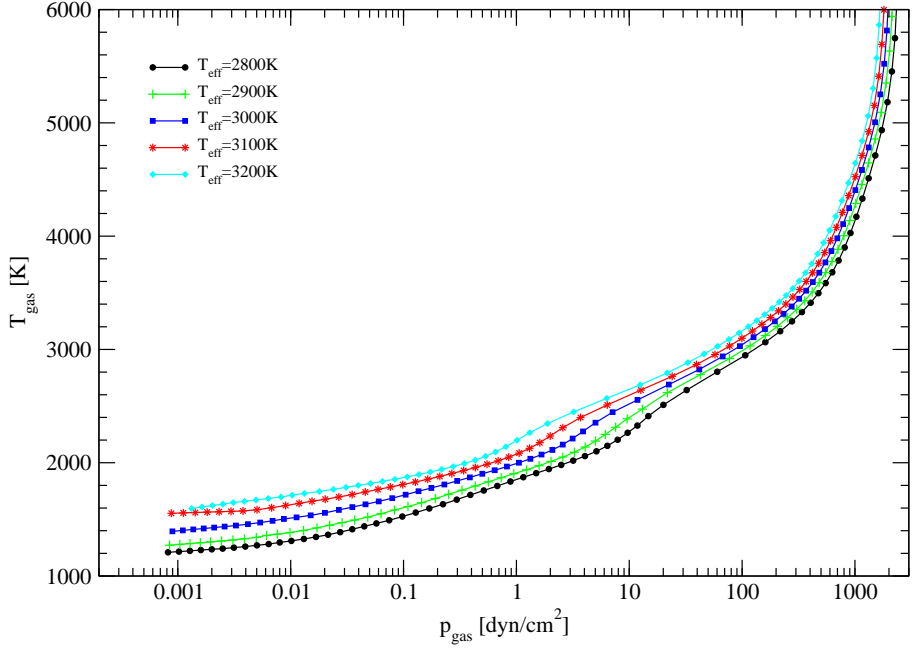}
\caption{Atmospheric structures of a sequence
of COMARCS models with $\rm log(g~[cm/s^2]) = 0.0$,
C/O = 1.4, $\rm Z/Z_{\odot} = 1.0$,
$\rm M/M_{\odot} = 1.0$ and different values of
$\rm T_{eff}$. The temperature is shown as a
function of the gas pressure.}
\label{aricphot01}
\end{figure*}

\begin{figure*}
\sidecaption
\includegraphics[width=12cm,clip]{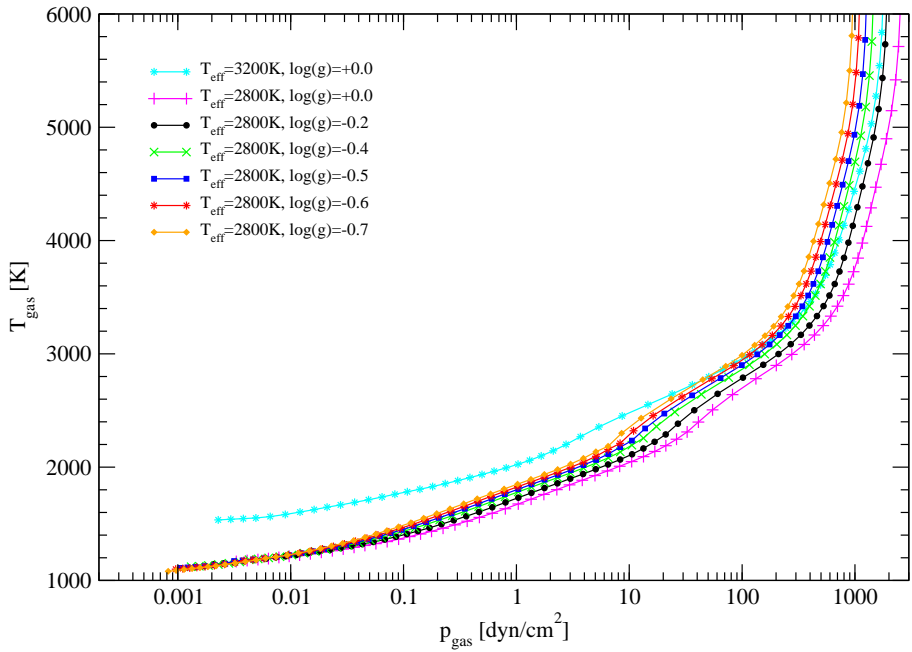}
\caption{Atmospheric structures of a sequence
of COMARCS models with C/O = 1.1,
$\rm Z/Z_{\odot} = 1.0$, $\rm M/M_{\odot} = 1.0$
and different values of $\rm log(g~[cm/s^2])$
at $\rm T_{eff} = 2800~K$. In addition, the
results for $\rm log(g~[cm/s^2]) = 0.0$
at $\rm T_{eff} = 3200~K$ have been included.
The temperature is shown as a function of
the gas pressure.}
\label{aricphot02}
\end{figure*}

The dominating molecules in our models of carbon stars
are CO, C$_2$, CN, C$_2$H$_2$, HCN and C$_3$. Their
contribution to the spectral absorption is shown in
Sect.~3.2 (Fig.~\ref{aricphot04}). Following the work of
Loidl et al.~(\cite{cphotloi01})
and the suggestions in the SCAN database
(J{\o}rgensen~\cite{cphotjor97})
we have scaled down the {\it gf}-values of the
C$_2$ lines taken from the original list of
Querci et al.~(\cite{cphotque74})
at wavelengths longer than 1.15~$\mu$m by a factor
of up to 10 in most of the calculations. This
has only a minor influence on the model structures,
but it causes moderate changes of the spectral
appearance in the region between 1.3 and 2.1~$\mu$m.
Thus, we have also produced some reference spectra
with unscaled C$_2$ data. The corresponding effects
on the photometric results will be discussed in more
detail later in this work.

As in the original version of COMA, Doppler profiles
including the thermal and the microturbulent contribution
are assumed for the molecules. First, there exists
almost no information on the damping constants of
molecular transitions. In addition, especially
for those species which dominate the overall opacity
like C$_2$H$_2$ or HCN, and in all regions close to
bandheads the wings of even the strongest lines
will be much weaker than the Doppler cores of the
many overlapping neighboring ones. The absorption
of all atoms except for hydrogen where we interpolate
in tabulated profiles is computed with full Voigt
functions, adopting the damping constants listed
in the VALD database.

\subsection{Model parameters}

The parameter range of the COMARCS atmospheres was chosen to
include the typical effective temperatures, surface
gravities and C/O ratios expected for carbon stars
from synthetic evolution calculations as they are
presented by Marigo \& Girardi~(\cite{cphotmar07})
or Marigo et al.~(\cite{cphotmar08}).
Sub-grids of models have been produced for the three
metallicities $\rm Z/Z_{\odot} = 1.0$,
$\rm Z/Z_{\odot} = 0.33$ and $\rm Z/Z_{\odot} = 0.1$
which covers the major populations in the Milky
Way as well as in the Magellanic Clouds. All elements heavier
than He were scaled with Z in the same way. Thus,
we did not take into account any possible individual
variations of the different species except for the
carbon abundance. However, due to the combination
of the fact that the opacities in cool giant atmospheres
are dominated by a small number of molecules with the
preferred formation of CO, the studied model structures
and photometric indices will mainly depend on [C] and
the ratio C/O\@. For the solar composition we adopted
the values from Anders \& Grevesse~(\cite{cphotand89})
except for C, N and O where we took the data from
Grevesse \& Sauval~(\cite{cphotgre94}). This is in
agreement with our previous work (e.g.\
Aringer et al.~\cite{cphotari99}) and results in a
$\rm Z_{\odot}$ of approximately 0.02.

All sub-grids were computed completely for objects with
a mass of $\rm M = 2.0~M_{\odot}$, which can be regarded
as typical for many carbon stars. At constant effective
temperature and surface gravity, this value determines
the ratio of the atmospheric extension to the stellar radius
and thus the overall spherical geometry. In order to
investigate the influence of the mass we also
produced a large number of models with
$\rm M = 1.0~M_{\odot}$. For solar metallicity,
almost the complete grid of 2.0~M$_{\odot}$
stars is covered by calculations for
1.0~M$_{\odot}$. Exceptions are some of the
most extended atmospheres where it was more
difficult to obtain converging hydrostatic
solutions for 1.0~M$_{\odot}$ in several cases,
as well as objects hotter than 3400~K which
are not on the AGB (see later in this
section). For the lower values of
$\rm Z/Z_{\odot}$, only a very small number
of 1.0~M$_{\odot}$ models restricted to an
effective temperature of 2600~K exists. This
is not a big problem, since the sphericity
corrections of the photometric indices in the
near infrared always remain relatively small,
as it can be concluded from Fig.~\ref{aricphot08}
and the corresponding discussion of the
results (Sect.~3.3.1). Uncertainties due to pulsation and
mass loss or unknown abundances are much
larger. In addition, some of the basic trends
of the metal-poor models as a function
of mass may also be deduced from the solar
metallicity atmospheres, although one should
keep in mind that the behaviour can sometimes
be quite complex. Finally, for
$\rm T_{eff} = 2800~K$,
$\rm log(g~[cm/s^2]) = -0.70$,
$\rm Z/Z_{\odot} = 1.0$ and C/O = 1.1
we have computed objects with 3.0, 5.0,
10.0 and 99.0~M$_{\odot}$ in order to
follow the convergence towards the plane
parallel solution.

For solar metallicity, our grid contains
C/O ratios of 1.05, 1.10, 1.40 and
2.00. Due to the much lower amount of oxygen,
the situation is quite different for
$\rm Z/Z_{\odot} = 0.33$ and
$\rm Z/Z_{\odot} = 0.1$. In such stars,
C/O values around 1.0 are expected to
appear rarely, since they may already
increase to much higher quantities after
one 'third dredge up' event. On the other hand, for
the more evolved carbon giants, ratios
considerably larger than 2.0 can be reached
(see Marigo \& Girardi~\cite{cphotmar07}). Thus,
for $\rm Z/Z_{\odot} = 0.33$ and
$\rm Z/Z_{\odot} = 0.1$ our grid includes
models with C/O = 1.4, 2.0 and 5.0.

The temperatures covered by the sub-grid for solar
metallicity range from 2400 to 4000~K with steps
of 100~K\@. For $\rm Z/Z_{\odot} = 0.33$ and
$\rm Z/Z_{\odot} = 0.1$ the lower limit was
increased to 2600~K, as expected from stellar evolution
calculations (e.g.\ Marigo et al.~\cite{cphotmar08}). The region
of AGB carbon stars extends only up to between 3200 and 3500~K,
depending on the initial mass and chemical composition.
Nevertheless, we also included hotter atmospheres, because the
computations predict some objects with C/O $> 1$
having a higher temperature during their post-AGB
phase. A typical sequence of models with
different values of $\rm T_{eff}$ at constant
surface gravity, mass and element abundances
is shown in Fig.~\ref{aricphot01}. In addition,
the effect of changing log(g) is demonstrated in
Fig.~\ref{aricphot02}.

Models with $\rm log(g~[cm/s^2]) = 0.0$
have been calculated for all combinations of
effective temperature, metallicity and
C/O ratio, except for some of the hotter
atmospheres between 3500 and 3800~K with a
high total carbon abundance
($\rm Z/Z_{\odot} = 0.33$ with C/O = 5.0
as well as $\rm Z/Z_{\odot} = 1.0$ with C/O =
2.0) where we could not obtain a converging
hydrostatic solution. This value of the surface
gravity is also the general upper limit of our
grid. It is obvious that for the cooler
objects it corresponds to luminosities
much lower than those expected for AGB
stars. For example, at $\rm T_{eff} = 2600~K$
and one solar mass, $\rm log(g~[cm/s^2]) = 0.0$
results in only 1122~L$_{\odot}$, which is by a
factor of 5 to 10 less than predicted for a
typical carbon giant. However, since it was
much easier to get converging solutions
for the higher surface gravities, we always
used this value of log(g) in order to iterate
from one grid temperature to the next.

\begin{table}
\begin{center}
\caption{Lower limits of log(g) in the grid of
atmospheric models for carbon stars. The upper
limit is always $\rm log(g~[cm/s^2]) = 0.0$.}
\begin{tabular}{cc}
\hline
$\rm T_{eff}$ range [K] & $\rm log(g~[cm/s^2])_{min}$\\
\hline
2400 --- 2700 & $-1.0$\\
2800 --- 2900 & $-0.9$\\
3000 & $-0.8$\\
3100 --- 3200 & $-0.6$\\
3300 --- 3400 & $-0.2$\\
3500 --- 4000 & $+0.0$\\
\hline
\end{tabular}
\label{aricphott1}
\end{center}
\end{table}

The lower limit for the surface gravity of our models
depends on the effective temperature. It was chosen
to cover the region of AGB carbon stars predicted by
Marigo \& Girardi~(\cite{cphotmar07})
and Marigo et al.~(\cite{cphotmar08})
and it is listed in Table~\ref{aricphott1}.
The smallest values of $\rm log(g~[cm/s^2]) = -1.0$
appear for the coolest giants with
$\rm T_{eff} \le 2700~K$, while they increase
continuously, when the objects become warmer.
In the range between 3500 and 4000~K we have only
computed atmospheres with $\rm log(g~[cm/s^2]) = 0.0$
and 2.0~M$_{\odot}$. As already mentioned
before, most or all of these stars will not be
on the AGB\@. The standard (maximum) step width in
$\rm log(g~[cm/s^2])$ is 0.2 above and 0.1 below
$-0.6$. However, due to difficulties concerning
the convergence of the more extended models
it had to be decreased in many cases by a factor
of two to 0.1 or 0.05. These problems are also
the reason why the lower limits given in
Table~\ref{aricphott1} could by far not be reached
with all combinations of metallicity, C/O ratio
and mass. Especially in situations with a high
carbon abundance ($\rm Z/Z_{\odot} = 0.33$
with C/O = 5.0, $\rm Z/Z_{\odot} = 1.0$
with C/O = 2.0) and around 3000~K, it became harder
(large number of iterations, highly dependent
on initial conditions) or impossible to obtain
hydrostatic solutions at decreased surface
gravities. In addition, it turned out
to be more difficult to calculate models with
a lower mass (1.0~M$_{\odot}$).

A list of all computed COMARCS models and synthetic spectra
including their parameters can be found in the bolometric
correction tables which are available at
{\tt http://stev.oapd.inaf.it/synphot/Cstars} or also at
the CDS\footnote{Data can be obtained via\\
http://cdsweb.u-strasbg.fr/cgi-bin/qcat?J/A+A/???/???}.

For the microturbulent velocity we adopted a value
of $\xi = 2.5$~km/s, which is in agreement with our
previous work (Aringer et al.~\cite{cphotari97})
and with high resolution observations of AGB stars
(e.g.\ Smith \& Lambert~\cite{cphotsmi90},
Lebzelter et al.~\cite{cphotleb08}).
Due to the fact that a large fraction of the
opacity in cool carbon giants is generated
by many weak overlapping lines, a change of
$\xi$ does only have a small effect on the
atmospheres and photometric results
as long as it remains moderate (e.g.\ between
2.0 and 3.5~km/s). The models were calculated
in the range from $\rm \tau_{Rosseland} = 10^2$
to $10^{-5}$ with a constant logarithmic
step size of 0.1. Thus, each of the structures
consists of 71 depth points. For a considerable
number of atmospheres we had to exclude a few of the
outermost layers in order to obtain converging
solutions. The missing data were added
subsequently by extrapolation. It was shown
with several test computations that the structures
produced by this method are usually close
to the ones determined considering the full depth
range. However, we never excluded more than
10 points (outer boundary of the calculation
at $\rm \tau_{Rosseland} = 10^{-4}$), since this
may easily result in larger errors, already visible
in the calculated spectra ($>1\%$).

\subsection{Synthetic spectra}

The described hydrostatic COMARCS atmospheres were
used to compute synthetic opacity sampling spectra
covering the range between 400 and 22500~cm$^{-1}$
(0.444 to 25.0~$\mu$m) with a resolution of
R = 10000. Based on the corresponding radial
temperature-pressure structures we derived the
necessary opacities in the different layers
again with the COMA code. Except for the
denser wavelength grid, this was done with the
same settings as for the generation of the input data
produced for the atmospheric models, ensuring a
consistent treatment of the absorption, which is
very important for obtaining realistic spectra
(see Aringer~\cite{cphotari05}). The results
of these opacity calculations are then processed
by a spherical radiative transfer program developed
for the work of Windsteig et al.~(\cite{cphotwin97})
that computes the desired stellar fluxes. This approach is
not completely consistent with the models, since COMARCS uses
a different set of routines. Nevertheless, for the
atmospheric parameters and wavelength ranges studied
here, the deviations between the low resolution opacity
sampling spectra generated directly by COMARCS and
the code from Windsteig et al.~(\cite{cphotwin97})
are negligible.

\begin{figure}
\centering
\includegraphics[width=8.5cm,clip]{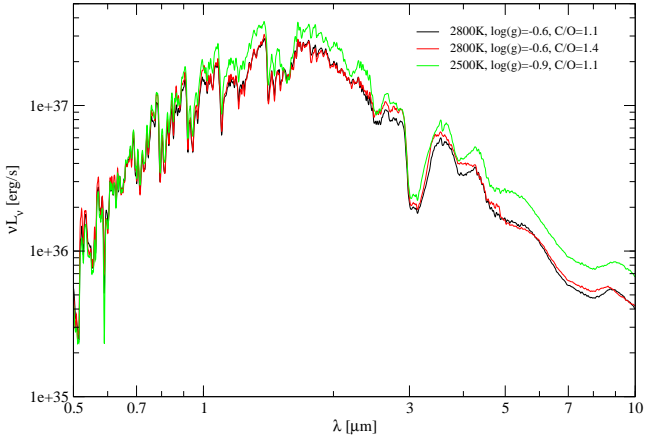}
\caption{Three spectra based on COMARCS models with
$\rm Z/Z_{\odot} = 1.0$, $\rm M/M_{\odot} = 1.0$
and different values of $\rm T_{eff}$, C/O and
$\rm log(g~[cm/s^2])$. The resolution is R = 200.}
\label{aricphot03}
\end{figure}

Due to the statistical nature of the opacity sampling
approach, only the average over a larger number of
wavelength points (usually 20 to 100) gives a realistic
representation of the observed stellar spectra. As a
consequence, we reduced the resolution of our results
to R = 200 by convolving them with a Gaussian function
defined by the corresponding half width. No additional
broadening caused for example by a macroturbulent
velocity was assumed. Some of the emerging spectra
are shown in Fig.~\ref{aricphot03} where the change
of the overall shape as well as of individual features
with effective temperature, surface gravity and
C/O ratio can be seen. It should be noted that
the actual sampling of the wavelength grid is with
R = 2000 significantly higher than the real resolution,
resulting in a relatively smooth appearance of the
plotted energy distributions. These R = 200 data
are electronically available as described in Sect.~3.2.

\subsection{Synthetic photometry}

Based on our theoretical results we have calculated synthetic
bolometric corrections following the same formalism as
described in Girardi et al.~(\cite{cphotgir02}, eq.\ 7 and 8).
For each of the different filters we first convolved the
total transmission curve with the original opacity sampling
spectra, integrating either the photon energy or the number
of detected photons. The latter selection depends on the
definition of the investigated photometric system, if amplifiers
or counting devices are used. Then, the obtained magnitudes
were scaled with respect to values corresponding to a reference
intensity in order to assess the zero points. This quantity
was determined from a convolution with the filter curve
applied to a constant flux per unit wavelength (frequency)
in the case of STmag (ABmag) systems or to a Vega ($\alpha$~Lyr)
spectrum for Vegamag systems. The standard star data were taken
from the work of Bohlin~(\cite{cphotboh07}). They are expected
to be accurate to within about 2\% in the optical and near
infrared range. Most of the photometric systems are defined
in a way that Vega has magnitudes close to zero in all filters
with just small corrections representing the offsets found
during observational campaigns such as, for example, the 2MASS
survey.

We deal with about 25 photometric systems, most of which
are listed in Table~2 of Marigo et al.~(\cite{cphotmar08})
together with their sources of transmission curves and reference
magnitudes. In addition, we recently inserted information on
the HST/WFC3 camera (Girardi et al.~\cite{cphotgir08}),
CFHT/Megacam (McLeod et al.~\cite{cphotmcl00},
Coupon et al.~\cite{cphotcou08}),
UT/DMC (Kuncarayakti et al.~\cite{cphotkun08})
and the Washington+DDO51 (KPNO Mosaic, Geisler~\cite{cphotgei84},
see also {\tt http://www.noao.edu/kpno/mosaic/filters})
filters. Since the number of entries in our database is continuously
growing, an updated overview may always be found at
{\tt http://stev.oapd.inaf.it/cmd}. Regarding the specific
case of carbon star spectra, we already include most of the
photometric systems in which these cool objects are more
interesting for observation, such as Bessell
JHK, 2MASS, UKIDSS, HST/NICMOS, OGLE, SDSS, AKARI and Spitzer.

\section{Results of the modelling}

\subsection{Model structures}

In Figs.~\ref{aricphot01} and \ref{aricphot02}
some typical examples for the hydrostatic models
computed with COMARCS are presented. One can see how
the temperature-pressure structure, which is of great
importance for the spectral appearance, changes as a
function of $\rm T_{eff}$ and log(g). It is obvious
that the variation of these two parameters produces
clear and uniform trends concerning the position
and shape of the corresponding curves. For example,
a decreased surface gravity will always result in
an atmosphere extending to lower densities at the
outer boundary, or with a growing value of
$\rm T_{eff}$ the whole structure is shifted to
warmer temperatures. Such a regular behaviour
usually also appears for the other parameters,
which are mass, metallicity and C/O ratio.
Thus, in many situations it will be possible to
determine atmospheres not included in the current
grid by interpolation or even extrapolation.
However, this has to be done with care, since
uniform clear trends do not occur in all cases
(see the discussion on the effect of mass
in the spectra in the following section, Sect.~3.2).

\subsection{Spectra}

The complete grid of our final spectra with R = 200 can be
obtained from {\tt http://stev.oapd.inaf.it/synphot/Cstars}.
The corresponding tables consist of three columns:
wavelength [{\AA}], continuum normalized flux and
specific luminosity times frequency ($\rm \nu \cdot L_{\nu}$
[erg/s]). The second quantity is determined from the
division of the result of a full radiative transfer
calculation including all opacities by one where
the atomic and molecular line absorption is completely
neglected (only continuum opacities). This allows us to evaluate
the apparent intensity of the different features.
As an example for the data we present in
Fig.~\ref{aricphot03} three spectra generated from
models with similar luminosity, but varying
effective temperature, surface gravity and C/O
ratio. They cover the interval from 0.5 to 10~$\mu$m,
which is only a part of the whole computed frequency
range extending between 0.444 and 25~$\mu$m. As it
has been mentioned in the previous section, the
resolution is R = 200, while the sampling of the
wavelength grid was chosen to be ten times higher.

\begin{figure*}
\sidecaption
\includegraphics[width=12cm,clip]{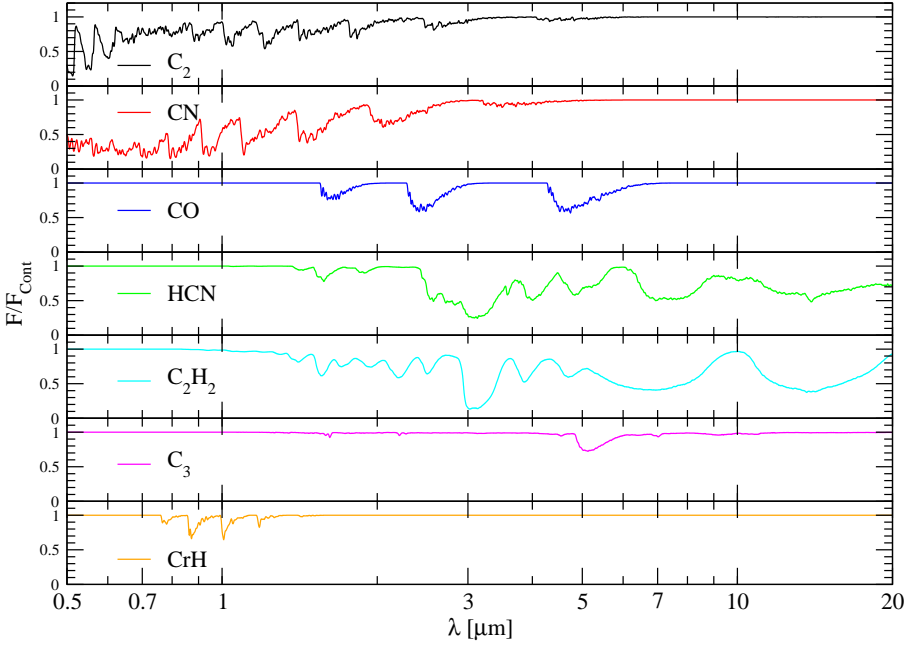}
\caption{Continuum normalized spectra for the most
important molecular species based on a COMARCS model
with $\rm T_{eff} = 2600~K$, $\rm log(g~[cm/s^2]) = -0.2$,
$\rm Z/Z_{\odot} = 1.0$, $\rm M/M_{\odot} = 1.0$ and
C/O = 1.10.}
\label{aricphot04}
\end{figure*}

In Fig.~\ref{aricphot04} we show spectra demonstrating the contribution
of the most important molecules, which have been derived by including
only the opacity of the corresponding species into the
radiative transfer. Subsequently, the results were
normalized to the continuum in the same way as described
above. We divided them by the output of a calculation where the absorption
generated by lines was completely neglected. All spectra are
based on the same atmospheric model with $\rm T_{eff} = 2600~K$,
$\rm log(g~[cm/s^2]) = -0.2$, $\rm Z/Z_{\odot} = 1.0$,
$\rm M/M_{\odot} = 1.0$ and C/O = 1.10. It is obvious
that molecular transitions block a large fraction of the
radiation in the shown frequency interval. The opacity
around and below 1~$\mu$m is dominated by the bands of
C$_2$ and CN, while at longer wavelengths C$_2$H$_2$ and
HCN are the most important species. However, the intensity
of the features generated by the latter polyatomic molecules
decreases significantly for higher effective temperatures.
On the other hand, the variation of the absorption produced
by C$_2$ and CN as a function of the stellar parameters is
much less pronounced. In addition, the bands of C$_3$ may become
considerably stronger with growing C/O ratios.

\begin{figure*}
\sidecaption
\includegraphics[width=12cm,clip]{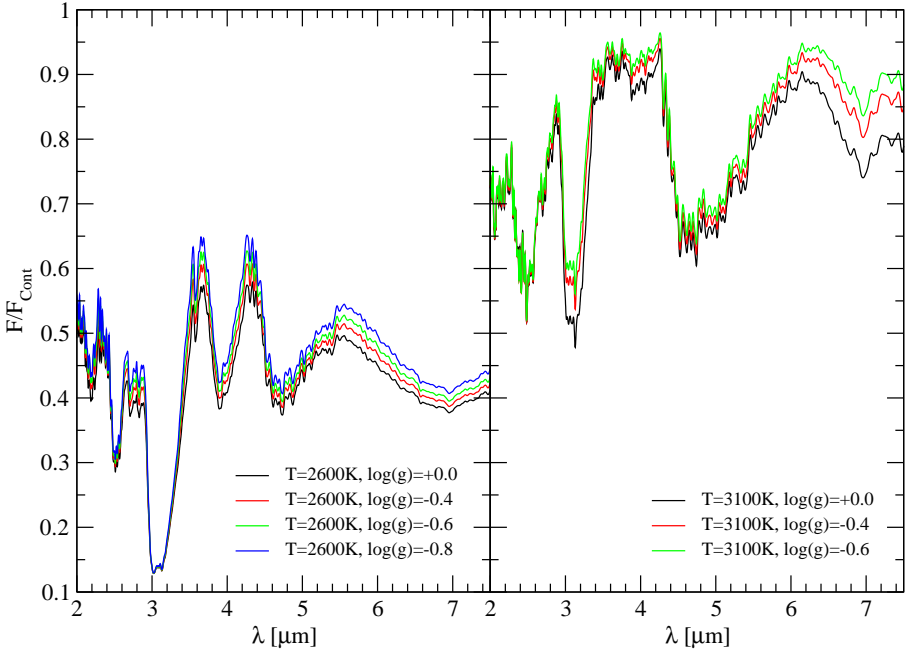}
\caption{Continuum normalized spectra based on COMARCS models
with $\rm Z/Z_{\odot} = 1.0$, $\rm M/M_{\odot} = 2.0$,
C/O = 1.10 and different values of $\rm log(g~[cm/s^2])$.
Two effective temperatures are compared: 2600~K and 3100~K.}
\label{aricphot05}
\end{figure*}

The behaviour of the molecular features depending on the stellar
properties of carbon stars has already been used
to determine parameters like effective temperature or
C/O ratio by a comparison of calculated and observed results
(e.g.\ Aoki et al.~\cite{cphotaok98}, Aoki et al.~\cite{cphotaok99},
J{\o}rgensen et al.~\cite{cphotjor00}, Loidl et al.~\cite{cphotloi01}).
Since this work focuses mainly on the photometric characteristics, we
will only discuss two selected examples of the spectroscopic changes,
which are interesting for a possible interpolation or extrapolation
based on our grid. More detailed studies involving the low resolution
energy distributions will follow in the future.

The change of the line absorption with surface gravity
is displayed in Fig.~\ref{aricphot05} where we show continuum
normalized spectra for different values of log(g) emerging
from cooler atmospheres with $\rm T_{eff} = 2600~K$ and hotter
ones with $\rm T_{eff} = 3100~K$. The models have solar
metallicity and their C/O ratio is 1.10. As one would
expect, the warmer object produces much weaker molecular
features. Even at the low resolution of the final spectra
the flux level of the continuum is almost reached at some
places around 4 and 6~$\mu$m. This is not the case for the
cooler atmosphere where the line absorption blocks in most
regions more than 50\% of the radiation. There exists
also a general trend that the molecular features become
slightly weaker at lower surface gravity, if the other
parameters are not altered. This behaviour is expected for
extended giants and caused by the larger dimensions of the
corresponding objects creating stronger sphericity
effects\footnote{In extended giants absorption lines or
bands will become weaker in a spherical atmosphere, since
they are filled with emission components originating
from the optically thin outer parts of the stellar
disk (e.g.\ Aringer~\cite{cphotari05}).} at constant
mass (Aringer et al.~\cite{cphotari99}). For both
of the presented sequences the changes as a function
of log(g) are reasonably linear. Since the variation
with temperature is usually also quite regular, an
interpolation concerning these two parameters should
work well.

\begin{figure*}
\sidecaption
\includegraphics[width=12cm,clip]{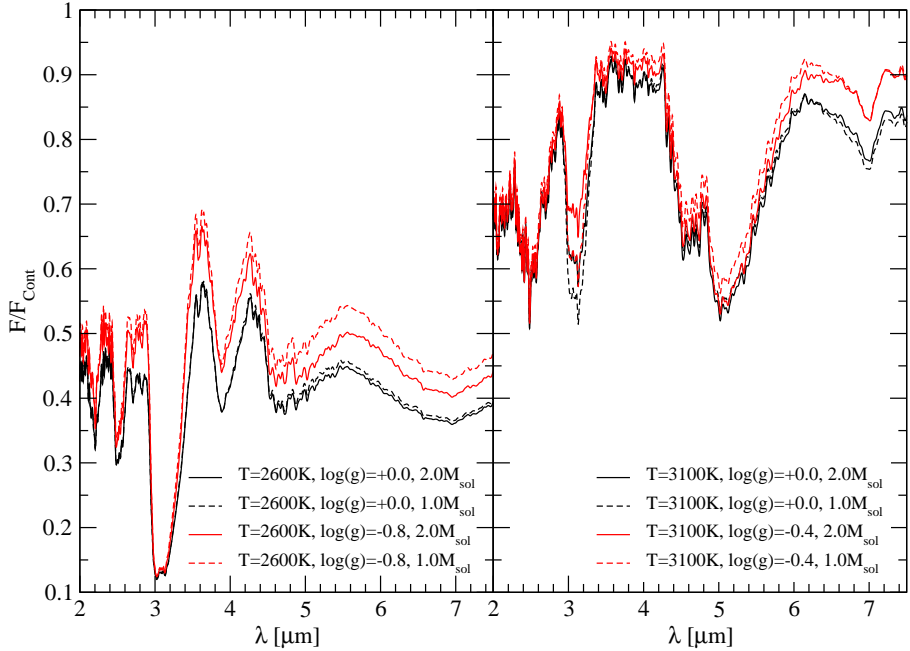}
\caption{Continuum normalized spectra based on COMARCS models
with $\rm Z/Z_{\odot} = 1.0$, C/O = 1.40 and different values
of $\rm log(g~[cm/s^2])$ and stellar mass (in M$_{\odot}$). Two
effective temperatures are compared: 2600~K and 3100~K.}
\label{aricphot06}
\end{figure*}

However, in some cases the situation can be more complex. This
is demonstrated in Fig.~\ref{aricphot06} where we investigate
the combined effects of mass and surface gravity at 2600 and 3100~K\@.
The spectra shown are again normalized with respect to the
continuum and the corresponding models have solar metallicity
and a C/O ratio of 1.40. At 2600~K we find a quite regular
behaviour which also complies with the expectations concerning
sphericity. As was discussed before (see Fig.~\ref{aricphot05})
the intensity of the line absorption decreases for lower values
of log(g). On the other hand, there is a shift due to mass
which grows for declining surface gravities. The molecular
features are in general deeper, if the objects become more massive.
Both trends are in agreement with the predicted weakening of the
line absorption due to sphericity effects. But at 3100~K
the behaviour is rather different. It is obvious that for this
temperature not all parts of the spectrum react in the same way
to changes of surface gravity and mass. Especially the deep
feature at 3~$\mu$m, produced mainly by HCN and C$_2$H$_2$,
does not really follow the trends expected from spericity,
since the intensity decreases a lot for more massive stars. A similar
behaviour can be found for the wings of the broad 14~$\mu$m
band, not shown in the plot. The absorption in the corresponding
region is also caused by the two species HCN and C$_2$H$_2$.
This demonstrates that due to the complex interaction of
atmospheric structure, opacities and chemical equilibrium,
simple predictions deduced from a single mechanism are not
always possible. On the other hand, there are spectral ranges where
we see also at 3100~K variations as a function of log(g) and
mass, which agree very well with the expectations from sphericity
(e.g.\ around 4 or 6~$\mu$m).

As will be discussed in the next section, the stellar mass
has in general only a rather small influence on the investigated
photometric properties. For the warmer stars with
$\rm T_{eff} \ge 3000~K$ the spectra below 2.5~$\mu$m also
change only marginally as a function of log(g). Thus, the
mentioned irregularities will usually not hamper a proper
prediction and interpolation of the colours. The possible
uncertainties due to circumstellar reddening by dust
and structural variations caused by atmospheric dynamics, which
are not considered in the presented models, are much larger,
even in objects with relatively weak pulsations. The special
behaviour of the 3~$\mu$m feature might affect the L band,
since it partly overlaps with the corresponding filter
curve.

\subsection{Photometry}

The synthetic photometry computed for the complete grid
of our hydrostatic COMARCS models is available at
{\tt http://stev.oapd.inaf.it/synphot/Cstars} or also at
the CDS\footnote{Data can be obtained via\\
http://cdsweb.u-strasbg.fr/cgi-bin/qcat?J/A+A/???/???}.
A detailed description of the corresponding data
can be found in Sect.~2.4. The tables list the
bolometric corrections (BC) as a function of the
stellar parameters $\rm T_{eff}$, $\rm log(g~[cm/s^2])$,
$\rm M/M_{\odot}$, $\rm Z/Z_{\odot}$ and C/O ratio
for the different included filter magnitudes. We have
prepared a separate file for each of the studied
photometric systems.

The complete set of tables covers 36 photometric
systems, each of them including several filters. This
wealth of information may be exploited by the users of the
database. Here, we restrict ourselves to certain typical
examples. We have chosen some of the standard visual
and near infrared filters as defined by
Bessell~(\cite{cphotbes90}) and
Bessell \& Brett~(\cite{cphotbes88}), since
they were applied for a large number of
observations. The V, J, H and K magnitudes
and colours appearing in the following discussions
and figures are based on this photometric
system (called the Bessell system in our work).

The observed colours of many carbon stars are severely affected
by circumstellar (see the discussion in Sect.~4) and interstellar
reddening. Since our results do not include these
processes, they have to be applied a posteriori to
the data.

\begin{figure}
\includegraphics[width=8.5cm,clip]{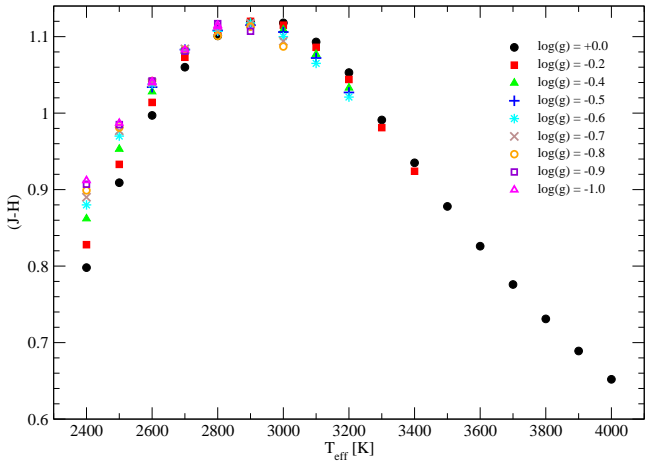}
\caption{Predicted (J$-$H) colours as a function
of effective temperature calculated from COMARCS models
with $\rm Z/Z_{\odot} = 1.0$, $\rm M/M_{\odot} = 2.0$,
C/O = 1.10 and different values of $\rm log(g~[cm/s^2])$}
\label{aricphot07}
\end{figure}

\subsubsection{The (J$-$H) colour}

\begin{figure}
\includegraphics[width=8.5cm,clip]{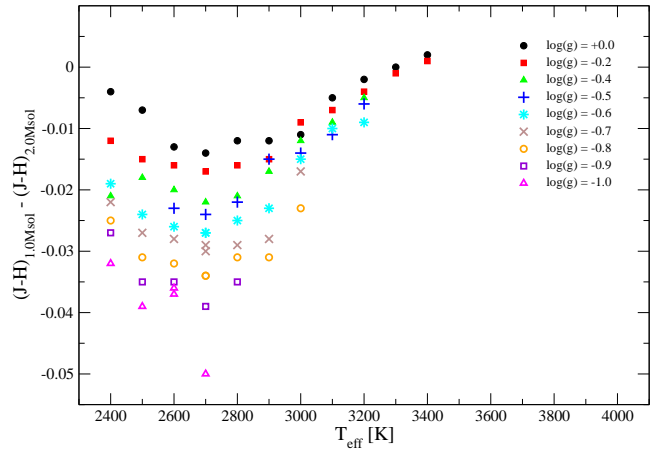}
\caption{Predicted differences in the (J$-$H) colours
between objects with 1.0 and 2.0~M$_{\odot}$ as a function
of effective temperature calculated from COMARCS models
with $\rm Z/Z_{\odot} = 1.0$, C/O = 1.10 and various
values of $\rm log(g~[cm/s^2])$}
\label{aricphot08}
\end{figure}

\begin{figure}
\includegraphics[width=8.5cm,clip]{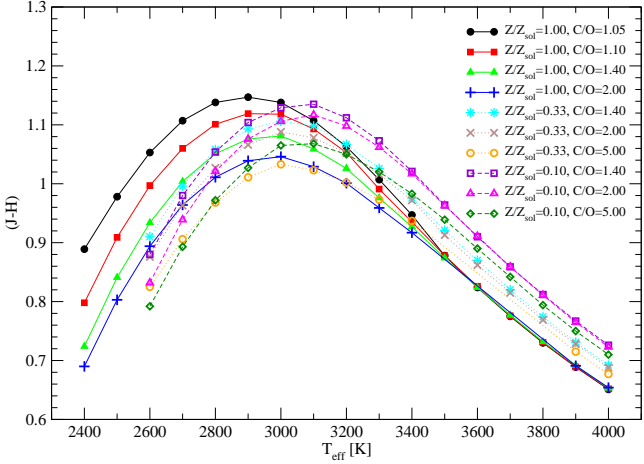}
\caption{Predicted (J$-$H) colours as a function
of effective temperature calculated from COMARCS models
with $\rm log(g~[cm/s^2]) = 0.0$, $\rm M/M_{\odot} = 2.0$
and different values of $\rm Z/Z_{\odot}$ (denoted
as $\rm Z/Z_{sol}$ in the plot) and C/O}
\label{aricphot09}
\end{figure}

In Figs.~\ref{aricphot07}, \ref{aricphot08} and
\ref{aricphot09} we investigate the behaviour of the
(J$-$H) colour as a function of the effective
temperature. In the first of the plots we do this
for the different values of log(g), selecting objects
with solar metallicity and C/O = 1.10 as an
example. As described in Sect.~2.2 (see
Table~\ref{aricphott1}), the range of available surface
gravities decreases towards warmer atmospheres and
above 3400~K only models with $\rm log(g~[cm/s^2]) = 0.0$
have been included in our grid. It is obvious that
the spread of the (J$-$H) colours has a significant
minimum around 2900~K, where it becomes practically
negligible. It grows quite rapidly, if the stars get
cooler. As a consequence, the largest differences
appear at the lowest temperature of 2400~K where the
dispersion reaches more than 0.1~mag. Above 2900~K
the spread also increases, but it remains always
moderate or small, which is partly due to the
limited log(g) range for the warmer atmospheres.
Below 2900~K the more extended models are redder, while
beyond this value they have bluer colours.

The value of 2900~K also corresponds to the point
of reversion for the general trend as a function of
temperature where (J$-$H) reaches a maximum. Towards
warmer and cooler atmospheres the colours become
bluer. This turnaround is caused by the behaviour
of different molecular features situated in the
region of the filters. In the J band the flux is
affected mainly by C$_2$ and CN, while the H band
covers a considerable depression due to a mixture
of C$_2$, CN, CO, C$_2$H$_2$ and HCN\@. The temperature
dependent intensity of the polyatomic species
especially plays a key role for the colour reversion,
which makes a determination of the stellar parameters
difficult and ambiguous. However, the turnaround
will usually not be visible in observations of real
stars, since all cooler objects are severely reddened
by dust.

The circumstellar reddening, which is discussed
in more detail in the next section (4.1) and in the
second paper of this series, affects all photometric
indices. It induces changes much larger than those
caused by the molecular features. Thus, the ambiguity
mentioned above plays only a minor role and for the
coolest stars the determination of the parameters
without a proper modelling of mass loss and dust
opacities is impossible.

In Fig.~\ref{aricphot08} we study the influence of the
stellar mass on the (J$-$H) colour at various values of the
surface gravity. The differences between 1.0 and 2.0~M$_{\odot}$
are shown taking again solar metallicity and C/O = 1.10
as an example. As one would expect, the shifts grow in general
for smaller values of log(g). In addition, they reach a maximum
around 2700~K\@. Their behaviour as a function of
temperature and surface gravity is not always completely
regular, which is in agreement with the results from the
synthetic spectra discussed in the previous section
(see Fig.~\ref{aricphot06}). A very good example is the
differences at $\rm log(g~[cm/s^2]) = -1.0$. Nevertheless, the
spread in (J$-$H) due to mass remains in most cases rather
small ($\le$~0.04~mag). Other effects, like changing the
elemental abundances or adding a dust opacity, are much
more important. The same is also
true for the other colour indices investigated here.

The temperature-dependent behaviour of (J$-$H) for
different chemical compositions is shown in
Fig.~\ref{aricphot09}. The sequences of models in the
plot represent various metallicities and C/O ratios
assuming $\rm log(g~[cm/s^2]) = 0.0$ and
$\rm M/M_{\odot} = 2.0$. The latter two values
have been chosen, since the corresponding COMARCS
atmospheres exist for most of the possible combinations
of the remaining parameters ($\rm T_{eff}$, C/O, Z).
It was already mentioned in the preceding discussion
that the trend of (J$-$H) with temperature reverses
at a certain point where the maximum of the index
is reached. Towards cooler and warmer models the colours
become bluer. In Fig.~\ref{aricphot09} one can see that this
applies to all investigated chemical compositions. However,
the maximum is shifted to higher temperatures, if the
metallicity becomes lower. At $\rm Z/Z_{\odot} = 0.33$
it is situated around 3000~K and at $\rm Z/Z_{\odot} = 0.1$
around 3100~K\@.

There exists a clear trend that in all atmospheres cooler
than about 3300 to 3400~K the (J$-$H) index becomes bluer towards
larger C/O ratios, if the metallicity and the other
stellar parameters are kept constant. The corresponding
spread in the colour temperature relation for our grid
reaches between 0.1 and 0.2~mag. The situation is quite
different for models that are warmer than 3400~K\@. In those
objects the value of C/O plays only a minor role. On the
other hand, one can see in Fig.~\ref{aricphot09} that at
higher temperatures the colours become redder, if the
metallicity decreases. A multiplication of $\rm Z/Z_{\odot}$
by a factor of one third gives a positive shift of 0.04 to
0.05~mag in (J$-$H). At a constant C/O ratio, which
introduces significant deviations below 3300 to 3400~K, this
rule applies down to about 3000~K\@. Around 2700 to 2800~K
the colour differences due to metallicity almost disappear.
And for the cooler temperatures we find a reversion of the
trend, because (J$-$H) slightly increases with growing
$\rm Z/Z_{\odot}$.

\begin{figure}
\includegraphics[width=8.5cm,clip]{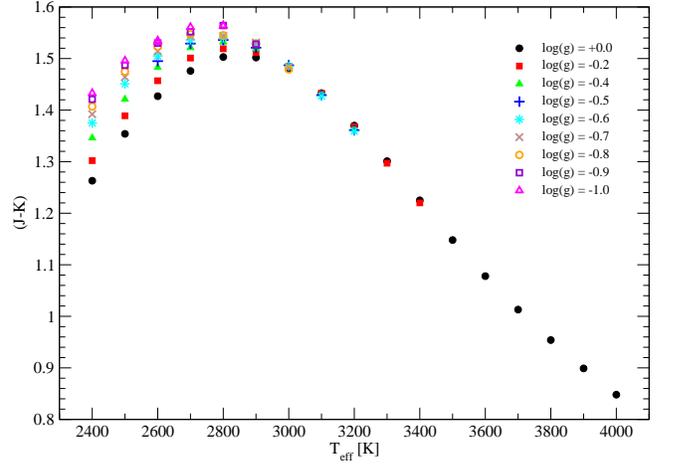}
\caption{Predicted (J$-$K) colours as a function
of effective temperature calculated from COMARCS models
with $\rm Z/Z_{\odot} = 1.0$, $\rm M/M_{\odot} = 2.0$,
C/O = 1.10 and different values of $\rm log(g~[cm/s^2])$}
\label{aricphot10}
\end{figure}

\subsubsection{The (J$-$K) colour}

\begin{figure}
\includegraphics[width=8.5cm,clip]{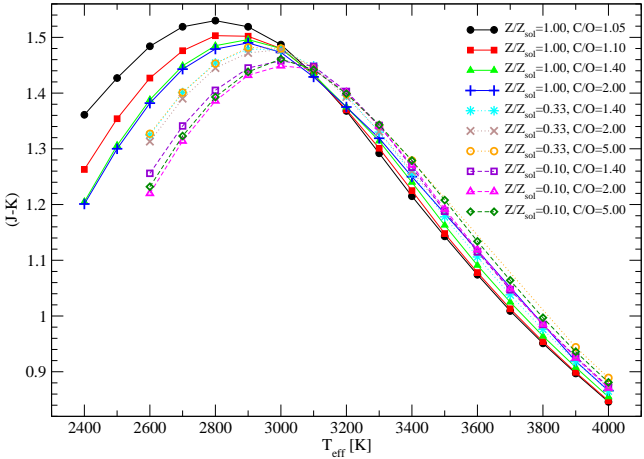}
\caption{Predicted (J$-$K) colours as a function
of effective temperature calculated from COMARCS models
with $\rm log(g~[cm/s^2]) = 0.0$, $\rm M/M_{\odot} = 2.0$
and different values of $\rm Z/Z_{\odot}$ (denoted
as $\rm Z/Z_{sol}$ in the plot) and C/O}
\label{aricphot11}
\end{figure}

The behaviour of (J$-$K) as a function of the effective
temperature is very interesting, since the corresponding
relation has often been used to determine that stellar parameter
from photometric observations. This may work quite well for the
warmer objects, as one can see in Fig.~\ref{aricphot10} where we
plot the colours of models with different values of log(g). Following
the previous discussion on (J$-$H), we have taken COMARCS
atmospheres with $\rm Z/Z_{\odot} = 1.0$, $\rm M/M_{\odot} = 2.0$
and C/O = 1.1 as an example. Above 2800~K the (J$-$K) index clearly
decreases towards higher temperatures. The scatter in the almost linear
relation, which is produced by the variation of the surface gravity in our
grid, remains negligible. From Fig.~\ref{aricphot11}, where we study the
effect of different chemical compositions for $\rm log(g~[cm/s^2]) = 0.0$
and $\rm M/M_{\odot} = 2.0$, it is evident that down to 3000~K
also changes of metallicity and C/O ratio do not create
too large deviations. Nevertheless, similar to the case of (J$-$H),
the trend with temperature reverses for the coolest models. As one
can see in Figs.~\ref{aricphot10} and \ref{aricphot11} the turnaround
with the reddest (J$-$K) appears at 2800~K for $\rm Z/Z_{\odot} = 1.0$
and a value of C/O below 1.4. It shifts to 2900 or even further to
3000~K, if the metallicity decreases and the relative abundance of
carbon becomes higher. The explanation for this reversion is similar
to the one for the behaviour of (J$-$H) in the coolest atmospheres.
The suppression of the flux in the K band due to molecular absorption
of polyatomic species grows at lower temperatures much more than in
the wavelength region where the J magnitude is determined. The K filter
covers mainly features of CN and C$_2$H$_2$. However, the turnaround
will usually also not become visible in populations of real stars, since
the coolest objects show a severe reddening caused by circumstellar
dust.

Looking at Fig.~\ref{aricphot10} it becomes obvious that
significant shifts introduced by the variation of the
surface gravity in our grid appear only below 2900~K\@.
The corresponding spread in the colour temperature relation
grows rapidly towards cooler models. At 2400~K the difference
between atmospheres with $\rm log(g~[cm/s^2]) = -1.0$ and
0.0 reaches almost 0.2~mag. In general, the more extended
objects are redder in (J$-$K). Fig.~\ref{aricphot11} shows
that below 3000~K the colour also depends on the chemical
composition. It gets bluer if the metallicity decreases.
For the coolest models a multiplication of $\rm Z/Z_{\odot}$
by a factor of one third reduces the index by approximately
0.05 to 0.06~mag. The C/O value also plays
a role below 3000~K\@. As long as the ratio remains small,
(J$-$K) decreases with higher relative carbon abundance.
However, this trend saturates, and above C/O = 1.4 there
are no noticeable differences. As already
mentioned, we find that the shifts due to the chemical
composition, which almost disappear between 3000 and
3200~K, never become large in the warmer models.
In general, at the higher temperatures
(J$-$K) turns slightly redder with an increasing C/O ratio,
while the behaviour as a function of metallicity is not
so regular and causes only minor deviations.

\begin{figure}
\includegraphics[width=8.5cm,clip]{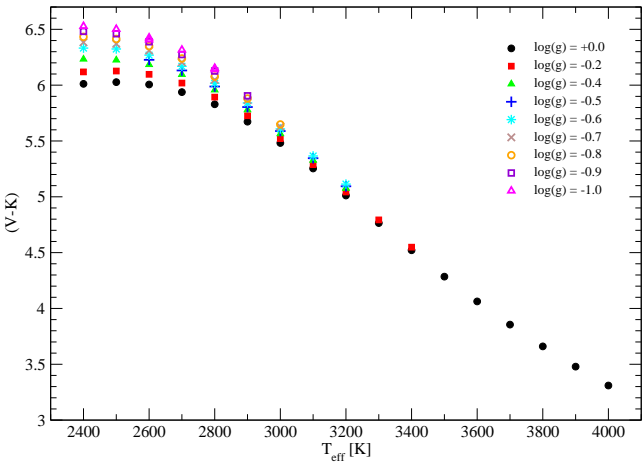}
\caption{Predicted (V$-$K) colours as a function
of effective temperature calculated from COMARCS models
with $\rm Z/Z_{\odot} = 1.0$, $\rm M/M_{\odot} = 2.0$,
C/O = 1.10 and different values of $\rm log(g~[cm/s^2])$}
\label{aricphot12}
\end{figure}

\subsubsection{The (V$-$K) colour}

\begin{figure}
\includegraphics[width=8.5cm,clip]{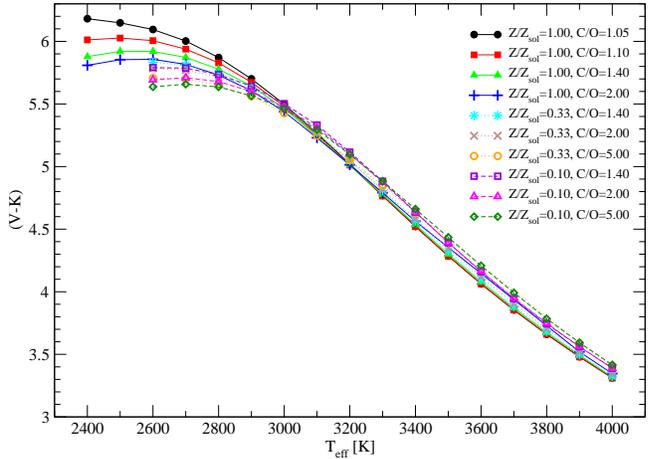}
\caption{Predicted (V$-$K) colours as a function
of effective temperature calculated from COMARCS models
with $\rm log(g~[cm/s^2]) = 0.0$, $\rm M/M_{\odot} = 2.0$
and different values of $\rm Z/Z_{\odot}$ (denoted
as $\rm Z/Z_{sol}$ in the plot) and C/O}
\label{aricphot13}
\end{figure}

As one can see in Figs.~\ref{aricphot12} and
\ref{aricphot13} the (V$-$K) index could be a quite good
indicator of the effective temperature. The relative scatter
due to variations of surface gravity and chemical abundances remains
in most cases reasonably small ($\rm < 0.2~mag \sim 100~K$).
Furthermore, in contrast to (J$-$H) or (J$-$K), the overall relation
that the colour becomes redder for cooler atmospheres is never reversed.
It only saturates at the lowest temperatures where the application of
hydrostatic dust-free models is not useful (see Sect.~4.1).
Unfortunately, there are almost no simultaneous measurements
of V and K magnitudes for individual carbon stars. In addition,
this index will be severely affected by interstellar as well as
circumstellar reddening.

In general, atmospheres with a smaller surface gravity
have redder (V$-$K) colours. In Fig.~\ref{aricphot12}
we demonstrate this for the carbon stars from our grid with
$\rm Z/Z_{\odot} = 1.0$, $\rm M/M_{\odot} = 2.0$ and
C/O = 1.10. One can see that significant shifts due to
the different values of log(g) appear only below 2900 to
3000~K, and they increase towards the coolest temperatures
where up to 0.5~mag may be reached. This applies
also to the variations caused by changes of the chemical
abundances. The behaviour of objects with
$\rm log(g~[cm/s^2]) = 0.0$ and $\rm M/M_{\odot} = 2.0$
is shown in Fig.~\ref{aricphot13}. The plot reveals
that at lower temperatures, models with a higher C/O
ratio and a decreased metallicity tend to produce bluer
colours. Above 3000~K this trend reverses, but the
corresponding relative differences do not become
considerable.

\section{Discussion}

\subsection{Comparison with observations}

In the previous section we showed some examples
representative of the results of our computations.
Now, we compare them to observational data. We will
restrict ourselves again to the Bessell photometric
system (Bessell~\cite{cphotbes90} and Bessell
\& Brett~\cite{cphotbes88}). We start with the work of
Bergeat et al.~(\cite{cphotber01}), who established
a calibration scheme for the effective temperature of
galactic carbon stars by relating interferometric results to
colours like (J$-$K), (H$-$K) or (V$-$K). The conversion of measured
angular diameters into stellar parameters depends on estimates
for the distance and functions describing the limb darkening
of the investigated objects. Especially if one studies
cooler AGB giants, both of these tasks may become quite
problematic. The difficulties are mainly caused by the
large extensions of the atmospheres, the diameters of which change
considerably with wavelength due to molecular absorption and with time
due to pulsation. These variations are often accompanied by complicated
radial density structures and the formation of dust shells
(e.g.\ Hofmann et al.~\cite{cphothof98}, Jacob \&
Scholz~\cite{cphotjac02}, Aringer et al.~\cite{cphotari08}).
Thus, the result will depend very much on the model applied
to interpret the intensity profile of the observed star
and the corresponding uncertainties. For such objects, even
the theoretical choice of a diameter to define the effective
temperature is rather arbitrary, since the range with an
optical depth around one may become quite broad.

\begin{figure}
\includegraphics[width=8.5cm,clip]{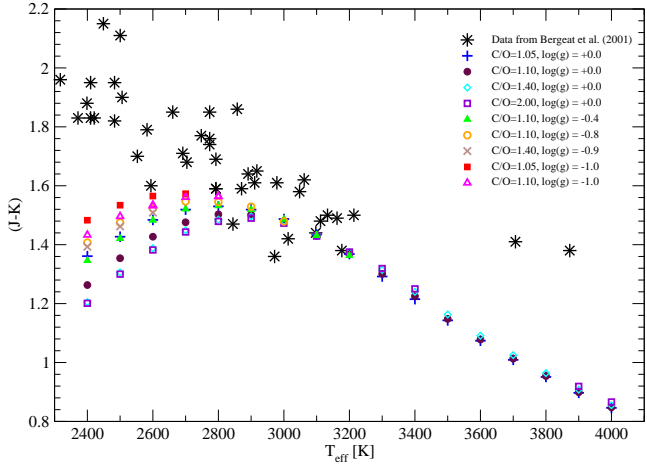}
\caption{The (J$-$K) colour as a function of the effective
temperature is shown for COMARCS models with
$\rm Z/Z_{\odot} = 1.0$ and $\rm M/M_{\odot} = 2.0$.
Several typical sequences characterized by different
values of $\rm log(g~[cm/s^2])$ and C/O are
included. The theoretical results are compared to the
data from Bergeat et al.~(\cite{cphotber01}, Table~4) which
have been derived from observations (stars).}
\label{aricphot14}
\end{figure}

\subsubsection{(J$-$K) versus effective temperature}

In Fig.~\ref{aricphot14}, where we show (J$-$K) as a function
of the effective temperature, COMARCS models with solar
metallicity are compared to the results of
Bergeat et al.~(\cite{cphotber01}). The sequences
of atmospheres with different values of log(g) and
C/O, which are included in the plot, were
selected in order to represent approximately the spread
introduced by the variation of these parameters
(see Figs.~\ref{aricphot10} and \ref{aricphot11}).
As one can see, the agreement between calculated and
observed colours is quite good in the range from
3200 down to 2800~K\@. The only distinction is that
the measured data show a much larger scatter than
the synthetic ones. However, this may very well be due
to uncertainties concerning the variability, the
correction of the interstellar reddening or the
determination of the linear diameters. Below 2800~K
the colours predicted by the COMARCS models are
systematically bluer than those taken from the
observations. While the (J$-$K) value in the results
of Bergeat et al.~(\cite{cphotber01}) continues to
increase towards the coolest stars, this trend
reverses for our hydrostatic calculations. As a
consequence, the difference between measured and
synthetic data grows considerably at lower
temperatures. Nevertheless, this behaviour is not
completely unexpected. We have already mentioned that
the cooler AGB giants are severely affected by
pulsation, dust formation and mass loss, which cannot
be described within the framework of hydrostatic
models (e.g.\ H\"ofner et al.~\cite{cphothof03}).
Due to circumstellar reddening and the development
of very extended and complicated temperature density
structures, these processes show a strong influence
on the spectra and filter magnitudes of the stars
(Gautschy-Loidl et al.~\cite{cphotgau04}).
The problem will be discussed in more detail
in the second paper of this series where we
present synthetic colours based on dynamical
calculations taking the corresponding time
dependent phenomena into account.
It is mainly the opacity of amorphous carbon
dust which changes the overall energy distribution
of the objects. It causes the high (J$-$K) values appearing
for the observed data in Fig.~\ref{aricphot14} and is not
included in our COMARCS models\footnote{An equilibrium
description of the dust in a hydrostatic atmosphere
results in much too high condensation degrees and
opacities.} which seem to be appropriate down to about
2800~K\@. We want to emphasize again that below this
point the conversion of measured angular diameters
into effective temperatures also will be problematic.

In Fig.~\ref{aricphot14} one can see two of the observed
stars situated at rather high temperatures above 3600~K
(BL~Ori, V4378~Sgr) which are far away from our COMARCS
sequences. However, as Bergeat et al.~(\cite{cphotber01})
already have noted they also do not fit into the relation
defined by the other measured objects. It is possible that
an error concerning the determination of the interferometric
data or their interpretation occured. In reality the stars
may be much cooler.

\begin{figure}
\includegraphics[width=8.5cm,clip]{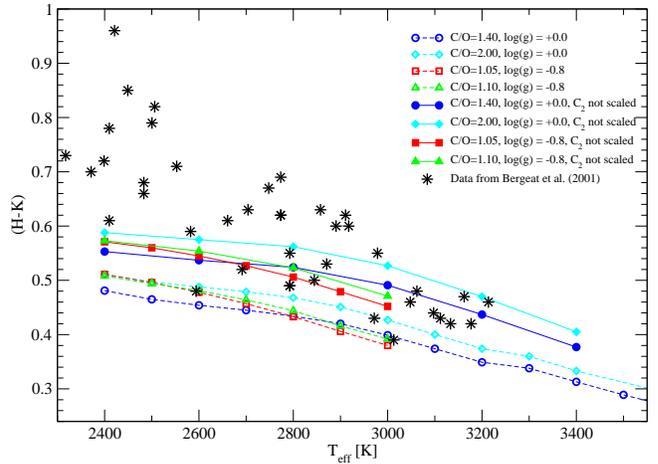}
\caption{The (H$-$K) colour as a function of the effective
temperature is shown for COMARCS models with
$\rm Z/Z_{\odot} = 1.0$ and $\rm M/M_{\odot} = 2.0$.
Several typical sequences characterized by different
values of $\rm log(g~[cm/s^2])$ and C/O are
included. Results from calculations with a scaled
C$_2$ line list (dashed lines, open symbols) and with the
unscaled original one (full lines, filled symbols) are
compared to each other and to the data from Bergeat
et al.~(\cite{cphotber01}, Table~4) which have been derived
from observations (stars).}
\label{aricphot15}
\end{figure}

\subsubsection{(H$-$K) versus effective temperature}

In Fig.~\ref{aricphot15} we present a comparison of the
observed colour temperature relation for carbon stars
with COMARCS models, which is similar to the one in
Fig.~\ref{aricphot14}, but shows (H$-$K) instead of
(J$-$K). The plot again includes sequences from our grid
of hydrostatic atmospheres with solar metallicity and
different values of log(g) and C/O (dashed lines,
open symbols). It is obvious that the (H$-$K) indices
produced by the calculations are systematically bluer
than the measured ones. Apart from the deviations below
2800~K growing towards cooler temperatures, which already
appeared for (J$-$K) and have been explained by the
contribution of circumstellar dust and dynamical changes
of the structures, there exists a shift of around
0.1~mag for all of the hotter objects. This
problem is not restricted to the data of
Bergeat et al.~(\cite{cphotber01}), but occurs
also in the two colour diagrams based on other
observations (not shown here).

In order to find possible explanations for the
differences between measured and predicted (H$-$K)
values of the warmer stars, we investigated the influence
of the scaling applied to the C$_2$ absorption in our
calculations. This has been proposed by
Loidl et al.~(\cite{cphotloi01}) and is described
in the section about model atmospheres and opacities
(Sect.~2.1). We produced a few series of synthetic spectra
and filter magnitudes using the original C$_2$ list
without any changes (full lines and filled symbols
in Fig.~\ref{aricphot15}). The relative
effect of the scaling on (H$-$K) is much stronger than on
the other Bessell colours discussed in our work. This
can be explained by the fact that the corresponding variations
reach a maximum in the H band and the comparatively small
flux difference between H and K\@. From looking at
Fig.~\ref{aricphot15} it becomes clear that the calculated
(H$-$K) indices based on the original linelist are by 0.08 to
0.1~mag higher, if the temperature is kept constant. The
consequence of the shift to redder colours is a much
better agreement with the observations in the region
above 2800~K, while below this limit we still see
considerable deviations, growing towards cooler stars.
Thus, the results for (H$-$K) with the unscaled C$_2$
opacity would confirm the ones derived from (J$-$K).
This may be interpreted as an indication that it is
preferable to use the original C$_2$ data without any
correction of the {\it gf}-values and to add approximately 0.1~mag
to the (H$-$K) indices from our standard grid. However, since
there is not much effect on the other colours, one should be
careful drawing conclusions concerning the molecular
absorption. Future work on the C$_2$ opacities and spectral
investigations can solve this problem.

\begin{figure}
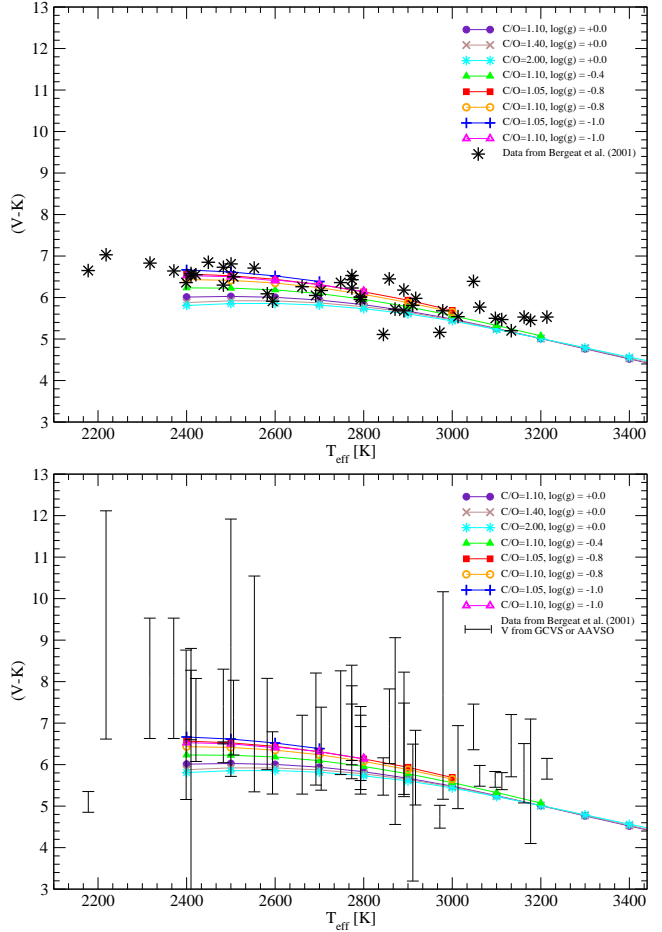

\includegraphics[width=8.5cm,clip]{aricphot16b.eps}
\includegraphics[width=8.5cm,clip]{aricphot16a.eps}
\caption{The (V$-$K) colour as a function of the effective
temperature is shown for COMARCS models with
$\rm Z/Z_{\odot} = 1.0$ and $\rm M/M_{\odot} = 2.0$.
Several typical sequences characterized by different
values of $\rm log(g~[cm/s^2])$ and C/O are
included. The theoretical results are compared to the
data from Bergeat et al.~(\cite{cphotber01}) which have
been derived from observations. In the upper panel we
present the original (V$-$K) indices given by the
authors (stars), while in the lower one we take the
V variability of the objects into account. The ranges
between the corresponding maxima and minima are shown
as error bars. The V and K magnitudes used for the second
plot were taken from various sources listed in the
text. They do not include the values from
Bergeat et al.~(\cite{cphotber01}).}
\label{aricphot16}
\end{figure}

\subsubsection{(V$-$K) versus effective temperature}

The last comparison of our solar metallicity COMARCS
sequences with the results of Bergeat et al.~(\cite{cphotber01})
is displayed in Fig.~\ref{aricphot16}, where we show (V$-$K)
as a function of the effective temperature. A plot based on the
original photometric data as they were published by the authors
can be seen in the upper panel. The obvious agreement between
predicted and observed colours in all regions of the diagram
is rather surprising, because from the preceding discussion
one would expect deviations for the cooler stars, which are due
to circumstellar reddening and dynamical changes of the
atmospheric structures. The influence of dust on (V$-$K)
should be much more pronounced than on the infrared indices
like (H$-$K) or (J$-$K). However, Bergeat et al.~(\cite{cphotber01})
have neglected the pulsation of the objects producing the strongest
variability at shorter wavelengths, as in the V range. Since
the measurements of the interferometric radii as well as of the
V and K magnitudes have not all been done simultaneously,
this may create some bias in the diagram.

In order to consider at least the V amplitude of the objects
studied by Bergeat et al.~(\cite{cphotber01}), we have determined
the (V$-$K) values corresponding to the maxima and minima of the
visual flux. The K magnitudes were assumed to remain constant.
The colour ranges obtained with this approach
are displayed in the lower panel of Fig.~\ref{aricphot16}
(bars). In most cases we could take the necessary photometric
data from the General Catalogue of Variable Stars
(GCVS4, Samus et al.~\cite{cphotgcv06}, Samus et al.~\cite{cphotgcv07}).
If the GCVS did not include the needed information, we
estimated the interval covered by the V magnitudes using the
AAVSO Light Curve Generator ({\tt http://www.aavso.org/data/lcg/}).
For a few of the objects we were not able to find any published
measurements revealing the temporal behaviour of the fluxes
in the V band (AB~Gem, V4378~Sgr, DR~Ser). Since all of them
are Lb (irregular) variables, which normally show rather moderate
photometric changes, we assumed the brightness listed in the General
Catalog of Cool Galactic Carbon Stars (GCCCS, Stephenson~\cite{cphotgcc89})
as the mean value and an amplitude of 0.5~mag. The K magnitudes
were taken from the $\rm 5^{th}$ edition
(Gezari et al.~\cite{cphotgez00}) of the Catalog of Infrared
Observations (Gezari et al.~\cite{cphotgez93}), the Two Micron All
Sky Survey (2MASS, Skrutskie et al.~\cite{cphot2mss06}), Whitelock et
al.~(\cite{cphotwhi06}) and Menzies et al.~(\cite{cphotmen06}).
For the correction of the interstellar reddening we used the
coefficients given by Bergeat et al.~(\cite{cphotber01}).

If one compares the upper and the lower panel of Fig.~\ref{aricphot16},
it becomes obvious that the original (V$-$K) data from
Bergeat et al.~(\cite{cphotber01}) are in most cases situated within the
range of variability, but not in its centre. They are systematically
shifted to bluer colours. The same is true for the values
predicted from our COMARCS models, which cover the same regions
of the diagram, as was mentioned before. All of this may be explained
by the fact that the authors have used V measurements biased towards
higher fluxes corresponding to phases of the pulsation where
the circumstellar extinction caused by dust remains weak.
However, also in the bottom panel of Fig.~\ref{aricphot16},
a clear increase in the differences between synthetic and
observed colours at lower temperatures is not visible.
Such a behaviour contradicts the trend found for (J$-$K) and
(H$-$K) as well as the expectations of stronger dynamical
effects and higher mass loss rates in cooler objects. Nevertheless,
since there are severe uncertainties concerning this diagram,
one should be rather careful with any definite conclusions. First,
we have neglected the variation of the K magnitudes and the angular
diameters, which is most likely less pronounced than the change of
the visual flux, but still quite considerable. In principle it would be
desirable to use simultaneous measurements or at least temporal
mean values of all involved quantities. Unfortunately, such data
are not available at the moment. Secondly, the applied corrections
of the interstellar reddening (Knapik \& Bergeat~\cite{cphotkna97})
may be problematic in the case of stars with intense pulsations,
because they are based on assumptions concerning a characteristic
energy distribution for the objects.

\begin{figure}
\includegraphics[width=8.5cm,clip]{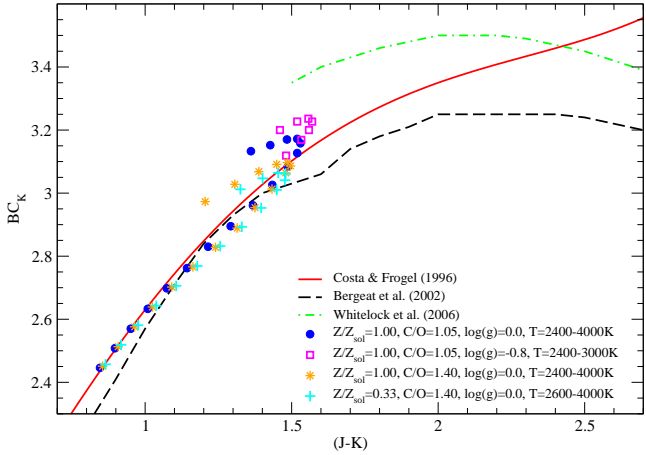}
\caption{The bolometric corrections for the K magnitude ($\rm BC_K$)
are shown as a function of the (J$-$K) colour. Several
representative sequences of COMARCS atmospheres characterized
by different values of $\rm log(g~[cm/s^2])$,
$\rm Z/Z_{\odot}$ (denoted as $\rm Z/Z_{sol}$ in the plot) and
C/O are displayed. The models, which cover the effective
temperature ranges given in the legend, have two solar
masses. The minima of $\rm BC_K$ correspond always to the
warmest object in each series. In addition, we have included
the observed mean relations for galactic carbon stars from
Bergeat et al.~(\cite{cphotber02}, dashed line),
Costa \& Frogel~(\cite{cphotcos96}, full line) and
Whitelock et al.~(\cite{cphotwhi06}, dotdashed line).}
\label{aricphot17}
\end{figure}

\subsubsection{Bolometric correction $\rm BC_K$}

In Fig.~\ref{aricphot17} we study the behaviour of the
bolometric corrections for the K magnitude ($\rm BC_K$)
as a function of the (J$-$K) colour. We compare several
effective temperature sequences of COMARCS atmospheres
with observed mean relations for galactic carbon stars
taken from the work of Bergeat et al.~(\cite{cphotber02}),
Costa \& Frogel~(\cite{cphotcos96}) and
Whitelock et al.~(\cite{cphotwhi06}). The included
model series correspond to different values of
metallicity, C/O ratio and surface gravity. Since
Costa \& Frogel~(\cite{cphotcos96}) claim that their
results may be used for LMC objects, we have also considered
a set of calculations assuming $\rm Z/Z_{\odot} = 0.33$.

From the hottest COMARCS atmospheres producing the smallest
(J$-$K) indices and bolometric corrections down to about
2800~K, the agreement between predictions and observations
is very good. In this range we see a clear relation that
with decreasing effective temperature (J$-$K) becomes redder
and $\rm BC_K$ larger. Nevertheless, at the cooler hydrostatic
models the trend reverses for both quantities and there are
deviations between the calculated positions in the diagram and the
measured mean sequences, which grow progressively. As was
discussed in the beginning of this section, the observed carbon
stars also extend to much higher (J$-$K) values than the COMARCS
atmospheres. The results of Whitelock et al.~(\cite{cphotwhi06}),
which are obviously focused on objects with considerable mass
loss rates, have almost no overlap with the computed
colours, since they only cover the interval down to
(J$-$K)~$\sim$~1.5. In agreement with the work of
Bergeat et al.~(\cite{cphotber02}), they reveal that for the
very red sources the bolometric corrections decrease again.
In the relation published by Costa \& Frogel~(\cite{cphotcos96})
such a behaviour does not appear. However, their results were mainly
determined for bluer carbon stars and are only extrapolated
towards larger (J$-$K) indices.

The most important conclusion from our plot of the bolometric
corrections versus (J$-$K) colour in Fig.~\ref{aricphot17}
is the confirmation of the scenario described
in the beginning of this section. Down to about 2800~K
the COMARCS models reproduce the observations quite
well, while at cooler temperatures the spectral flux
distributions of the real stars are dominated by circumstellar
dust shells and dynamical changes of the atmospheric
structure.

\begin{figure}
\includegraphics[width=8.5cm,clip]{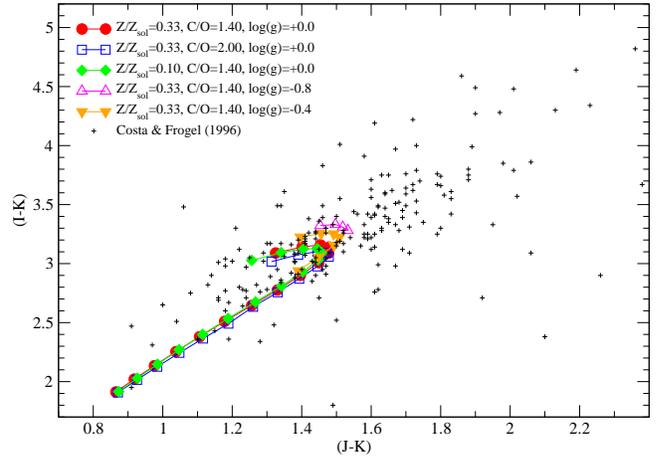}
\caption{Two colour diagram (I$-$K) versus (J$-$K) for
several representative sequences of COMARCS atmospheres
characterized by different values of $\rm log(g~[cm/s^2])$,
$\rm Z/Z_{\odot}$ (denoted as $\rm Z/Z_{sol}$ in the plot) and
C/O\@. All models have two solar masses. The series start
at $\rm T_{eff} = 2600~K$ and range up to 4000~K for
$\rm log(g~[cm/s^2]) = 0.0$, up to 3200~K for
$\rm log(g~[cm/s^2]) = -0.4$ and up to 2900~K for
$\rm log(g~[cm/s^2]) = -0.8$. The warmest objects
correspond to the bluest colours. The data are compared
to observations of LMC carbon stars published by
Costa \& Frogel~(\cite{cphotcos96}).}
\label{aricphot18}
\end{figure}

\subsubsection{Near infrared two colour diagram}

Another confirmation of this scenario can be found in
Fig.~\ref{aricphot18}, where we show a two colour diagram
with (I$-$K) versus (J$-$K). Several representative temperature
sequences of subsolar metallicity COMARCS atmospheres
are compared to the observations of LMC carbon stars published by
Costa \& Frogel~(\cite{cphotcos96}). Again, the included
model series correspond to different values of log(g),
$\rm Z/Z_{\odot}$ and C/O\@. It is obvious that for the
warmer objects having bluer colours the agreement between
predicted and measured positions in the diagram can be
regarded as good. On the other hand, as in
Fig.~\ref{aricphot17}, the observed data extend to
much larger (I$-$K) or (J$-$K) indices than the
calculations. They also do not show any signs of the
reversion of the trend with temperature appearing
for the COMARCS atmospheres below 2800~K\@. A similar
behaviour confirming the discussed scenario can also be found
in most of the other two colour diagrams combining different
Bessell filters (not shown here).

Samples of distant carbon stars are often selected due to the
red colours of these objects. A good example is the criterion
of $\rm (J - K) > 1.4$ (e.g.\ Cioni \& Habing~\cite{cphotcio03}).
However, from the predicted and observed values in
Fig.~\ref{aricphot18} it becomes clear that such a
photometric choice will neglect many of the bluer
sources with low mass loss rates.

\subsection{Application of the data to stellar evolution}

\subsubsection{Defining an interpolation scheme}

The models presented in this paper define a 5-dimensional
grid in metallicity, effective temperature, surface gravity,
C/O ratio and mass. Nevertheless, the latter has a
quite small influence on the synthetic colours and bolometric
corrections, as it is shown in Fig.~\ref{aricphot08}. In addition,
the atmospheres for many combinations of the other parameters
have only been calculated with 2.0~M$_{\odot}$. A more extensive
grid involving 1.0~M$_{\odot}$ computations exists only for
solar metallicity. As a consequence, the effect of stellar mass
may be treated as a correction to be applied a posteriori to
the results, and we remain with a 4-dimensional interpolation
to determine the desired photometric quantities.

For an object with a given value of Z, $\rm T_{eff}$, log(g) and
C/O we first identify the metallicity interval $\rm (Z_1,Z_2)$
of the grid where it is located and compute the weighting factors
for a linear interpolation in log(Z). Subsequently, we look for
the proper temperature intervals $\rm (T_{eff,1},T_{eff,2})$ at
Z$_1$ and Z$_2$. These are not necessarily equal, since due to
problems with the convergence of the more extended models
and the different expected properties of carbon stars
(see Sect.~2.2), the spacing and range of coverage with respect to
a certain parameter may change as a function of the other
stellar quantities. For example, at solar metallicity our
calculations go down to 2400~K, while for $\rm Z/Z_{\odot} = 0.33$
and $\rm Z/Z_{\odot} = 0.1$ the lower limit was increased to 2600~K
following the predictions from synthetic AGB evolution. From each
$\rm T_{eff,1}$ and $\rm T_{eff,2}$ we are then able to determine
the weighting factors for a linear interpolation in
$\rm log(T_{eff})$. As a next step we use the same method
for log(g) where the interval (log(g)$_1$, log(g)$_2$) has to be
defined for all combinations of Z$_1$ and Z$_2$ with
$\rm T_{eff,1}$ and $\rm T_{eff,2}$. Finally, we include also the
C/O ratio into this scheme. In the ideal case, when the
investigated object is situated completely inside all limits,
we remain with 16 final weighting factors
computed from the product of the four individual
contributions of the mentioned parameters which may then
be applied to the colours or bolometric corrections of the
corresponding grid points. Extrapolations are in general
avoided by selecting the available maximum or minimum
values. They were only allowed for the effective temperature
with an amount of up to 100~K\@.

\begin{figure}
\includegraphics[width=8.5cm,clip]{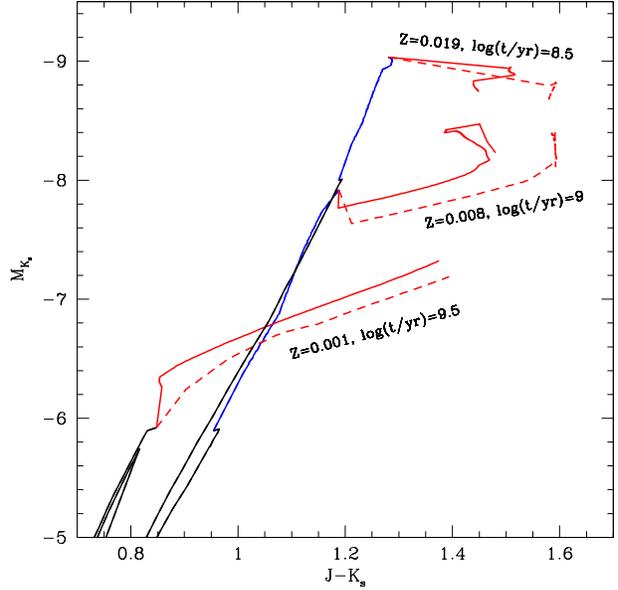}
\caption{Selected isochrones from Marigo et al.~(\cite{cphotmar08}),
focusing on the part of the 2MASS $\rm M_{K_s}$ versus $\rm (J-K_s)$
diagram that corresponds to the upper RGB (black lines) and TP-AGB phases.
To avoid confusion in the figure the TP-AGB part covers only the
quiescent stages, neglecting the thermal pulse cycle variations, and we
do not show the final points in which stars cross back to the blue
during their evolution to the post-AGB\@. The TP-AGB is marked either
as oxygen rich (blue lines in the electronic version) or as carbon rich
(red lines). Full lines are results obtained using the present
database of transformations for carbon stars, whereas dashed lines
are the ones from Marigo et al.~(\cite{cphotmar08}) based on spectra
from Loidl et al.~(\cite{cphotloi01}). The effect of circumstellar
dust is not considered. The ages ($\rm log(t~[yr])$) and metallicities of
the isochrones are indicated in the plot.}
\label{aricphot19}
\end{figure}

\subsubsection{Theoretical isochrones}

Interpolated bolometric corrections computed in this way can then be
applied to theoretical TP-AGB models. Fig.~\ref{aricphot19} illustrates
the difference between the results based on our data to convert the
isochrones of Marigo et al.~(\cite{cphotmar08}) to the 2MASS
$\rm M_{K_s}$ versus $\rm (J-K_s)$ diagram and the original ones obtained
with the spectra taken from Loidl et al.~(\cite{cphotloi01}). The latter
were only available for a few selected stellar parameters (mainly
temperatures) and solar metallicity. We have chosen curves with
$\rm log(t~[yr]) = 8.5$ for $\rm Z = Z_{\odot} = 0.019$, $\rm log(t~[yr]) = 9.0$
for $\rm Z = 0.42~Z_{\odot} = 0.008$ and $\rm log(t~[yr]) = 9.5$ for
$\rm Z = 0.053~Z_{\odot} = 0.001$, because the typical age range for the
appearance of the carbon star branch changes with the chemical
abundances. The effect of circumstellar dust has been neglected
in the plot. As it was already mentioned it would shift
the objects with mass loss to much redder colours.

The figure shows that at all metallicities the AGB C-stars are
predicted to be distributed along a sort of tail redward of the
location of the oxygen rich giants. This behaviour is largely due
to lower effective temperatures caused by the opacity of carbonic
molecules as has been described in Marigo~(\cite{cphotmar02}),
Marigo et al.~(\cite{cphotmar03}) and Marigo \& Girardi~(\cite{cphotmar07}).
Applying the atmospheric models presented here to the determination
of photometric properties results in some changes in the discussed
colour magnitude diagram. Compared to the calculations based on the spectra
of Loidl et al.~(\cite{cphotloi01}) the objects become bluer and in most
cases brighter. As a consequence, especially at the two higher metallicities,
the maximum $\rm (J-K_s)$ value decreases by up to 0.15~mag. A part of the
flux differences can be attributed to a variation of the photometric
zero points caused by the more limited wavelength coverage of the
data from Loidl et al.~(\cite{cphotloi01}). The bulk of the change
is due to revised opacities and the much larger parameter range included
in the current grid. As an example, Loidl et al.~(\cite{cphotloi01})
did not consider any atomic lines and only solar metallicity spectra were
available for the conversion of the isochrones.

Observed carbon star red tails in systems like the Magellanic Clouds
present the majority of giants extending more or less uniformly in the
interval $\rm 1.2 \la (J-K_s) \la 2.0$. The results shown in
Fig.~\ref{aricphot19} do not reproduce the redder part of this
distribution. As it has been discussed, this is mainly due to the
neglect of the circumstellar dusty envelopes. Also, the effect
of the pulsation on the atmospheric structure and colour excursions
caused by thermal pulse cycles may play a role. Evaluating the impact
of such phenomena remains beyond the scope of the present paper. Work
is in progress to create a complete simulation of the Magellanic
Cloud TP-AGB stars considering all the processes affecting their
photometric properties.

The present database of bolometric corrections has already been
incorporated into the interactive web interface
{\tt http://stev.oapd.inaf.it/cmd} which can be used to generate
interpolated Marigo et al.~(\cite{cphotmar08}) isochrones and
their derivatives in many photometric systems. The tables containing
the BC values as well as the synthetic spectra are provided in the
repository {\tt http://stev.oapd.inaf.it/synphot/Cstars}.
The reddening caused by circumstellar dust may currently
be simulated in connection with our data by applying
the approximative approaches from Bressan et al.~(\cite{cphotbre98}) and
Groenewegen~(\cite{cphotgro06}). This is the same procedure
as used for the results of Marigo et al.~(\cite{cphotmar08}).

\section{Conclusions}

We have produced a grid of hydrostatic COMARCS atmospheres covering
effective temperatures between 2400 and 4000~K, surface gravities
from $\rm log(g~[cm/s^2]) = 0.0$ to $-1.0$, metallicities ranging
from the solar value down to one tenth of it and C/O ratios
in the interval between 1.05 and 5.0. Based on these models we
calculated synthetic low resolution spectra and bolometric corrections
for a considerable number of standard photometric systems, which are
publicly available on the web. As an example we have shown some of the
Bessell colours as a function of the stellar parameters. It turned
out that the mass, which represents in principle the sphericity
of the atmospheres, usually has only a quite small effect on the overall
energy distribution. On the other hand, as one would expect, the effective
temperature is the most important quantity. Especially for the warmer
carbon stars its determination based on photometric measurements may
reveal rather reliable results, if one chooses the right colour indices
and has at least a rough estimate of the other parameters.
In addition, the effect of the interstellar reddening has to be
taken into account, which represents the main source of uncertainties,
at least for individual stars.

However, we have also demonstrated that our hydrostatic dust-free
atmospheres fail to reproduce the redder and cooler carbon stars.
The photometric properties based on the models from our grid
were compared to several observed data sets involving effective
temperatures obtained from interferometry, bolometric corrections
and two colour diagrams. Most of these investigations confirm the
same scenario. Down to about 2800~K the agreement between predicted
and measured energy distributions is quite good, while below this
limit the COMARCS atmospheres are much too blue. In some cases
the photometric indices of the coolest models even decrease again
due to the influence of molecular features. Such a behaviour
definitely does not appear for the observed carbon stars, which
always extend to considerably redder colours than any of the
calculations. The main explanation for these differences is the
neglect of the dusty circumstellar envelopes in the presented
COMARCS atmospheres. But dynamical changes of the radial
pressure temperature structure connected to pulsation and mass
loss also play an important role.

Many time-dependent phenomena appearing in AGB stars, like
the intense pulsation creating shock waves or dust formation
driving heavy stellar winds, cannot be described within the
framework of hydrostatic atmospheres in chemical equilibrium, as it
has been demonstrated by a large number of models and observations
(e.g.\ Aringer et al.~\cite{cphotari99}, Alvarez \& Plez~\cite{cphotalv98},
Loidl et al.~\cite{cphotloi99}, H\"ofner et al.~\cite{cphothof03}).
Thus, a forthcoming paper of this series will focus on the effect
of dynamics and mass loss on the photometric properties of
carbon rich giants. There we demonstrate that these processes
offer a natural explanation for the very red colours seen in
a large number of objects. The calculations will be based on models
including pulsation and time-dependent dust formation as have
been published by H\"ofner et al.~(\cite{cphothof03}) or
Mattsson et al.~(\cite{cphotmat08}).

For the warmer COMARCS models, which are in agreement with the
observations, we found that the photometric changes as a function
of the stellar parameters show in most cases clear and predictable
trends. An exception is the dependence on mass, which corresponds
to the effect of sphericity. However, as we have demonstrated,
the latter plays only a minor role for the different colours.
Thus, it is possible to interpolate between the photometric
fluxes based on our grid, which allows us to connect the results
to stellar evolution calculations. The effect of applying
our bolometric corrections to the isochrones from
Marigo et al.~(\cite{cphotmar08}) has been shown in this
paper as a first test. For the moment we have neglected
the influence of dynamical processes and the reddening caused
by the dusty envelopes of the stars. The latter
will be included in a more systematic study involving
stellar evolution and population synthesis computations that
will be presented in a forthcoming publication. It can
already be taken into account in combination with our data
by using approximative descriptions like the ones from
Bressan et al.~(\cite{cphotbre98}) and Groenewegen~(\cite{cphotgro06}).

\begin{acknowledgements}
We acknowledge financial support from the University of Padova
(Progetto di Ricerca di Ateneo CPDA052212). LG acknowledges support
from PRIN INAF07 1.06.10.03. The work presented here was supported
by the Austrian Science Fund (FWF) projects P19503-N16 and P18939-N16.
BA acknowledges funding by the contract ASI-INAF I/016/07/0.
MTL has been supported by the Austrian Academy of Sciences (DOC
programme) and acknowledges funding by the Austrian Sience Fund
(FWF) project P-18171. We thank U.G. J{\o}rgensen and R. Gautschy-Loidl
for their support concerning the opacity data of C$_3$ and C$_2$H$_2$.
We acknowledge with thanks the variable star observations from the
AAVSO International Database contributed by observers worldwide and
used in this research. We thank A. Bressan and M. Groenewegen
for their early interest in this work and for many useful suggestions.
\end{acknowledgements}

\end{document}